\documentclass[sigconf,balance=false, nonacm]{acmart}

\usepackage{amsmath}
\usepackage{amsthm}

\theoremstyle{definition}

\newtheorem{lemma}{Lemma}

\usepackage{pifont} 

\usepackage{cleveref}

\usepackage{listings}
\usepackage{xspace}

\usepackage{enumitem}

\usepackage{msc-custom}

\usepackage{tikz}
\usepackage{pgfplots}
\pgfplotsset{compat=1.18}
\usepgfplotslibrary{statistics}
\usetikzlibrary{arrows.meta, positioning}

\usepackage[operators]{cryptocode}

\usetikzlibrary{positioning,fit,arrows.meta,backgrounds,shadows.blur}
\usetikzlibrary{shapes.multipart}

\usepackage[colorinlistoftodos,prependcaption,textsize=small]{todonotes}

\usepackage{hyperref}
\hypersetup{
    colorlinks = true,
    linkcolor = blue,
    anchorcolor = blue,
    citecolor = blue,
    filecolor = blue,
    urlcolor = blue
}

\usepackage[caption=false,font=footnotesize]{subfig}

\begin{document}

\newcommand{\eg}{\emph{e.g.,}\xspace}

\newcommand\encircleB[1]{%
	\tikz[baseline=(X.base)]
	\node (X) [draw, shape=circle, inner sep=.5, fill=blue, text=white, minimum size=.05cm, scale=0.8] {\textbf{#1}};%
}
\newcommand\encircleW[1]{%
	\tikz[baseline=(X.base)]
	\node (X) [draw, shape=rectangle, inner sep=1.5, fill=green, text=black, minimum size=.05cm, scale=0.8] {\textbf{#1}};%
}


\newcommand{\para}[1]{\par\noindent\emph{#1.}} 

\definecolor{darkpastelgreen}{rgb}{0.01, 0.75, 0.24}
\definecolor{darkpastelred}{rgb}{0.76, 0.23, 0.13}

\newcommand{\cmark}{\textcolor{darkpastelgreen}{\ding{51}}}
\newcommand{\xmark}{\textcolor{darkpastelred}{\ding{55}}}

\newcommand{\procs}{\mathcal{P}}
\newcommand{\pin}{\mathsf{in}}
\newcommand{\pout}{\mathsf{out}}
\newcommand{\pelse}{\mathsf{else}\ }
\newcommand{\plet}{\mathsf{let}\ }
\newcommand{\pthen}{\mathsf{then}\ }
\newcommand{\ppair}{\mathsf{pair}}
\newcommand{\punp}{\mathsf{unpair}}
\newcommand{\plook}{\mathsf{lookup}}
\newcommand{\pn}{\mathsf{new}\ }
\newcommand{\pset}{\mathsf{set}}
\newcommand{\pget}{\mathsf{get}}
\newcommand{\pins}{\mathsf{insert}}
\newcommand{\st}{\text{ such that }}
\newcommand{\fst}{\mathsf{fst}}
\newcommand{\snd}{\mathsf{snd}}
\newcommand{\encr}{\mathsf{aead\_enc}}
\newcommand{\decr}{\mathsf{aead\_dec}}
\newcommand{\KDF}{\mathsf{KDF}}
\newcommand{\pdiscard}{\mathtt{DISCARD}}
\newcommand{\repet}{\text{\textexclamdown}}

\newcommand{\hash}{\mathsf{h}}

\title{Association-based Privacy Attacks in Wireless Protocols: Formal Modeling and Mitigation}

\author{Mohit Kumar Jangid}
\orcid{0009-0000-1563-6974}
\affiliation{%
   \institution{Indian Institute of Technology Jodhpur}
   \city{Jodhpur}
   \state{Rajasthan}
   \country{India}}
\email{mjangid@iitj.ac.in }

\author{Felix Engelmann}
\orcid{0000-0001-9356-0231}
\affiliation{%
   \institution{Ohio State University}
   \city{Columbus}
   \state{Ohio}
   \country{USA}}
\email{fe-research@nlogn.org}

\author{Zhiqiang Lin}
\orcid{0000-0001-6527-5994}
\affiliation{%
   \institution{Ohio State University}
   \city{Columbus}
   \state{Ohio}
   \country{USA}}
\email{zlin@cse.ohio-state.edu}

\renewcommand{\shortauthors}{Jangid et al.}

\keywords{User privacy, exclusive-use, network protocols, IoT, deanonymization, user profiling, tracking, formalization, privacy risks}


\begin{abstract}
With the surge in privacy-sensitive data from sources such as social media and IoT devices, there is a pressing need for formal, automated methods to assess privacy risks within these intricate systems. This paper formally investigates root sources of pairing-based privacy threats exploited using replay/relay techniques in wireless communication. Our research harnesses condition-oblivious responses, replay-resistance, and  distance bounding measures vital for protocols utilizing shared keys in allowlists for authenticated reconnections. Particularly, the paper uses formal modeling of notable wireless networks, like the \mbox{Wi-Fi} P2P persistent group formation and the Bluetooth Low Energy reconnection procedure, to illustrate the root causes and countermeasures. Our model rigorously validates the proposed solution against association inference attacks, along with existing formalizations of well-authentication, frame opacity, and no-desynchronization. The ensuing analysis reveals not only uncharted privacy realms in wireless communication but also identifies old and new vulnerabilities. Our proposed design changes are acknowledged by \mbox{Wi-Fi} Alliance and Bluetooth SIG, paving the way for future advancements in resilient, privacy-preserving wireless protocols.

\end{abstract}

	\maketitle



\section{Introduction}
\label{sec:introduction}

In the digital era, individuals and organizations rely heavily on technologies for instantaneous data exchange. Notably, businesses amass and analyze vast volumes of user data, leveraging its insights to refine their services and maintain a competitive edge in the market~\cite{gordon2013big}. However, this continuous flow of information underscores the paramount importance of privacy, particularly in sectors such as the Internet of Things (IoT), healthcare~\cite{dash2019big,tawalbeh2020iot}, and intelligence/government agencies. Furthermore, breaches of privacy for public figures can have a profound impact on their reputation and professional life~\cite{TheHuman64:online}. Given these stakes, ensuring privacy has become an urgent imperative, necessitating robust mechanisms to guarantee the safety and security of sensitive information.

As privacy is recognized as granting consent and control over various aspects of one's life, including communication, intellectual pursuits, physical presence, and well-being, the privacy property \emph{unlinkability} has been studied extensively in the domain of digital communication~\cite{privacy-terminologies-2008, privacy-terminologies-2016, privacy-2-condition-proverif-2016, privacy-3-conditions-tamarin-2020, unlinkability-research-4, unlinkability-research-5}. Unlinkability is achieved when an external observer monitoring multiple users' actions (such as communication via their devices) cannot correlate two distinct communications to a single user. If an adversary succeeds to link both, they could deduce intimate details about an individual's interests, location, and health status, among other things, compromising the user's anonymity. Notably, unlinkability has been studied in a range of protocols: e-voting~\cite{evoting-selene-tamarin-2017,evoting-belenios-tamarin-2019}, TLS~\cite{kemtls-tamarin-2022}, backhaul network~\cite{trackability-tamarin-ndss-2023}, 5G~\cite{trackability-tamarin-ndss-2023,privacy-3-conditions-tamarin-2020}, e-passports ~\cite{bac-protocol-9303-2006, pace-protocol,privacy-3-conditions-tamarin-2020}, \mbox{Wi-Fi}~\cite{ssid-deanonymize-I-2016, ssid-deanonymize-II-2015}, etc.

In the domain of wireless communication,  especially IoT applications, various unlinkability threats have emerged in the past. These smart devices employ a peculiar feature known as Preferred Network List (PNL) to connect frequently connecting devices efficiently. Trading off usability for security, this feature has been exploited to de-anonymize large crowds based on preferred \mbox{Wi-Fi} SSIDs list~\cite{ssid-deanonymize-I-2016, ssid-deanonymize-II-2015, ssid-deanonymize-III-2018}.  Advancing the impact further, the Bluetooth MAC Address Tracking Attack  (BAT Attack)~\cite{bat-attack-exclusive-use-2022} deanonymizes users at an individual level. In this attack, the attacker leverages \emph{allowlists} (the PNL for Bluetooth devices) and replay/relay techniques to track Bluetooth user locations. These threats reveal that wireless protocols may harbor subtle vulnerabilities leading to privacy breaches. The nature of these privacy threats is further complicated by the fact that different contexts of communication mediums, authentication mechanisms, and data share architectures impact unlinkability differently. Given these challenges, we notice a lack of foundational root causes of PNL-based attacks. Therefore, we undertake a formal investigation of privacy attacks exposing the association of users.
In doing so, we ask the following questions: \emph{How can we systematically investigate the fundamental causes of these breaches? What countermeasures are effective at neutralizing these threats, both theoretically and in practice?}  \looseness=-1

To address these questions, this paper conducts a comprehensive, formal investigation of the mechanisms underlying allowlist-based attacks. To tackle the challenges of formal modeling and reasoning about unlinkability, we expand existing formal building blocks: Hirschi et al.~\cite{privacy-2-condition-proverif-2016} and Baelde et al.~\cite{privacy-3-conditions-tamarin-2020} use a combination of trace properties and diff-equivalence to reason about unlinkability based on three sufficient conditions (\S\ref{sec:background-related}): Well-Authentication (WA), Frame Opacity (FO), and No-Desynchronisation (ND).
 We instrument privacy using the state-of-the-art symbolic verification tool Tamarin ~\cite{tamarin-white-paper}, preferring it over DeepSec~\cite{deepsec-tool-2018}, APTE~\cite{apte-tool-2014} or AKISS~\cite{akiss-tool-2016} because Tamarin can reason about unbounded processes, sessions and messages providing stronger and generic guarantees.

In this research, we find that the presence of allowlists in wireless protocols and conditional checks over this data are the root causes enabling an adversary to deduce user \emph{association} in privacy-sensitive groups. For example, BAT attacks capitalize on locally shared data such as allowlists, commonly set up among IoT devices within organizational or personal frameworks for swift reconnections~\cite{bat-attack-exclusive-use-2022}. This means that an adversary with the ability to relay or replay packets between a distant device associated with a targeted user's privacy-sensitive group, whether it be a household or paired phone, can disclose the user's association and even pinpoint their location. It is worth noting that in wireless communications, relay or replay can be easily and cost-effectively achieved due to the ubiquity of smartphones and programmable software-defined radios, as opposed to wired communications, as shown in various digital contact tracing works~\cite{ellis2022replay,avitabile2022privacy,casagrande2021contact}.

Modeling allowlists in a process calculus reveals attacks that were not captured in previous work.
By using Tamarin~\cite{tamarin-white-paper}, we formally model Bluetooth reconnection procedures and \mbox{Wi-Fi} P2P persistent group formation. Through detailed analysis, we highlight design flaws in the protocols leading to Association Inference (\emph{AInf}) and propose corrective countermeasures. We employ a combination of condition oblivious responses, replay-resistant techniques, and distance bounding checks to propose design changes mitigating AInf attacks. Our Tamarin models confirm the WA, FO, and ND privacy properties, in addition to the Replay-Resistant (RR) measures and Distance Bounding (DiB)\cite{distance-bounding-herirachy-2018}). Additionally, to demonstrate the preliminary performance overhead of our proposed design, we develop C++ protocol codes for \mbox{Wi-Fi} and Bluetooth reconnection procedures. The evaluation indicates that the proposed protocol design achieves reconnection within 80-125 milliseconds as long as distance bounding checks can be optimized within reasonable bounds.  Encouragingly, both \mbox{Wi-Fi} Alliance and Bluetooth SIG have acknowledged our findings and agreed with our proposed solutions. Furthermore, both interest groups plan to publish our reports to stakeholders.
We make the following contributions:

\textbf{Novel Privacy Characterization
(\S\ref{sec:new-privacy-vector-formalization}).} We systematically characterize AInf privacy attack, expanding our understanding beyond existing privacy formulations. \looseness=-1

\textbf{Revised Protocols Design(\S\ref{sec:new-privacy-vector-formalization}).} Our recommendations for protocol design modifications specifically target the prevention of AInf, concurred by both Bluetooth SIG and \mbox{Wi-Fi} Alliance. Our proposed design strategically combines distance bounding checks and condition oblivious responses to deterministically detect relay and replays while keeping the faster reconnection feature intact.

\textbf{Formally and Practically Validated Case Studies (\S\ref{sec:case-studies}).} By formally verifying the proposed protocols for two prevalent real-world wireless protocols—Bluetooth and \mbox{Wi-Fi} P2P—we have demonstrated the generality and robustness of our privacy formalization approach. Additionally, we developed a C++ protocol code providing preliminary performance overhead of the proposed techniques.

\section{Background \& Related Works}
\label{sec:background-related}

\para{Bluetooth Pairing} Bluetooth is a short-range wireless technology that transitioned from simple device pairing to facilitating mesh networks for streamlined IoT operations. During the initial connection, devices pair using a comprehensive authentication procedure~\cite{bt-spec-5.2}, which consists of two distinct authentication phases and human interaction, resulting in the sharing of a Long-Term Key (LTK), such as the Identity Resolution Key (IRK).
Thereafter, the reconnection process begins by discovering devices using advertisements of Random Private Addresses (RPA)~\cite{bat-attack-exclusive-use-2022}. These addresses are composed of randomized hashed values of IRK shared at the end of the first-time pairing protocol.
A receiving device then searches its local storage for registered IRKs to identify the sender. Upon mutual RPA recognition, an ephemeral session key is derived for subsequent encrypted data exchanges~\cite{4.2-general}.

\para{\mbox{Wi-Fi} P2P or \mbox{Wi-Fi} Direct~\cite{wifi-direct-spec-1.9}} WiFi P2P/Direct facilitates peer-to-peer communication, enhancing IoT device interaction by bypassing Access Points (AP). This direct approach saves energy and boosts network adaptability and robustness. In this setup, devices dynamically assume access point or client roles and utilize the \mbox{Wi-Fi} Protected Setup (WPS)~\cite{wps-spec-2.0.8} for security. It supports group formations for various applications, such as home automation and vehicle-to-vehicle communication. The \mbox{Wi-Fi} P2P specification categorizes devices as either \emph{Supplicants}: requesting authentication, or \emph{Authenticators}: granting access. Group connections can be initiated internally or externally and form in three ways: standard, autonomous, and persistent. While standard and autonomous connections dynamically designate device roles and fully authenticate new members, persistent groups utilize stored network IDs and associated Pre-Shared Keys (PSK) for faster reconnections.  \looseness=-1

\para{BAT Attack~\cite{bat-attack-exclusive-use-2022}} BAT attack exploits Bluetooth's MAC addresses to track users. Even with MAC address randomization, BLE devices periodically broadcast their MAC address advertisements, allowing attackers to correlate devices. The vulnerability becomes more pronouncd when BLE devices enable the \emph{allowlist}, responding only to a authorized \emph{preferred} devices. Attackers can track these devices by analyzing MAC addresses and responses.
An active form of this attack can be carried out by replaying a sniffed MAC address, probing whether a peripheral or central device responds. This exploitation is facilitated by a flaw in the current MAC address randomization algorithm of the Bluetooth specification, which does not specify a method for random number selection and lacks safeguards against the reuse of randomness. As a result, attackers can track devices across the randomization time interval by replaying MAC addresses and assessing whether they are in the allowlist. The BAT attack, therefore, leverages both replay and relay attack vectors to exploit user privacy and track locations, with the relay attack being independent of the replay attack. \looseness=-1

\para{Related Works} Previous research has explored the use of human-readable SSIDs in \mbox{Wi-Fi} and Bluetooth for deanonymizing locations and interests of individuals~\cite{bat-attack-exclusive-use-2022} and groups~\cite{ssid-deanonymize-I-2016, ssid-deanonymize-II-2015, ssid-deanonymize-III-2018}. Unlike these studies, our work goes beyond the advertisement/probe phase and comprehensively characterizes the allowlist privacy attack. Prior formal research on  privacy~\cite{privacy-2-condition-proverif-2016,privacy-3-conditions-tamarin-2020} focuses on defining building block privacy properties with case studies on RFID and 5G networks. In contrast, our work leverages these building block properties and expands the attack scope to cover relay attacks over uncovered case studies for \mbox{Wi-Fi} and Bluetooth with mitigations.

The closest work to ours is IDBleed~\cite{idbleed-ndss-chris-2025}, which adopts a measurement approach to explain the widespread occurrence of exclusive-use attacks without formalization. It offers a defense using channel broadcasting and requiring synchronization. Instead, our focus lies on a formal approach to precisely specify Bluetooth and \mbox{Wi-Fi} protocol sections vulnerable to WA, ND, and FO privacy properties. Moreover, our defense does not require synchronization, and we propose a concrete protocol design that defends against both replay and relay attacks. \looseness=-1

\begin{figure}[t]
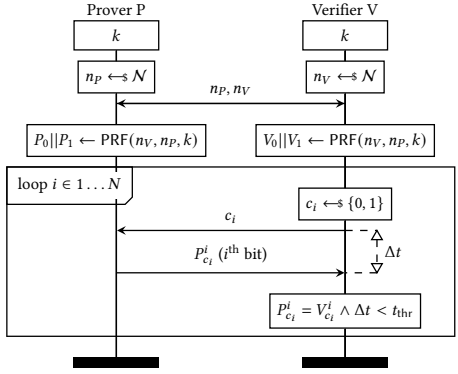
 
	\centering
	\vspace{-5mm}
	\scalebox{0.7}{
	\begin{msc}[instance distance=2.7cm, draw frame=none, msc keyword={},title top distance=0pt,title distance=0pt,
	first level height=2mm, level height=4mm]{}
\declinst{p}{Prover P}{$k$}
\declinst{v}{Verifier V}{$k$}
\action*{$n_P\sample\mathcal{N}$}{p}
\action*{$n_V\sample\mathcal{N}$}{v}
\nextlevel[2]
\mess{$n_P, n_V$}{p}{v}
\mess{}{v}{p}
\nextlevel
\action*{$P_0||P_1\gets\mathsf{PRF}(n_V,n_P,k)$}{p}
\action*{$V_0||V_1\gets\mathsf{PRF}(n_V,n_P,k)$}{v}
\nextlevel[2]
\inlinestart[right inline overlap=20.7mm,left inline overlap=20.7mm]{loop}{\colorbox{white}{loop $i\in 1\dots N$}}{p}{v}
\nextlevel
\action*{$c_i\sample\{0,1\}$}{v}
\nextlevel[2]
\measure[side=right]{$\Delta t$}{v}{v}[2]
\mess{$c_i$}{v}{p}
\nextlevel[2]
\mess{$P_{c_i}^i$ ($i$\textsuperscript{th} bit)}{p}{v}
\nextlevel
\action*{$P_{c_i}^{i}=V_{c_i}^{i}\land \Delta t<t_\mathsf{thr}$}{v}
\nextlevel[2]
\inlineend{loop}

\end{msc}}
\vspace{-6mm}
	\caption{Distance bounding by Hancke and Kuhn~\cite{hancke-kuhn-distance-bounding-2005}.}\label{fig:distance-bound-hanckle-kuhn}
\end{figure}

\para{Distance Bounding Protocol~\cite{distance-bounding-white-paper-1994}} Distance Bounding Protocol was first introduced by Brands and Chaum in 1994 to protect against relay attacks that can compromise a wide range of systems, including touchless payment card transactions, wireless automobile keys, and RFID access controls~
\cite{distance-bounding-herirachy-2018, dib-practical-1-2015, dib-practical-2-2012, dib-survey-2021}. Relay attacks occur when an adversary acts as a man-in-the-middle, relaying authenticated connection packets from the source to the target device. Since relaying bypasses cryptographic protections, detecting the proximity of peer devices through distance bounding is critical.
Distance Bounding Protocols establish the average response time and ensure it remains within an application-specific threshold through a series of timed challenge-responses.
The verifying device uses the speed of the electromagnetic wave (speed of light), as well as the response time from the peer device, to calculate proximity. Given that this is a time-sensitive calculation, multiple challenge responses are exchanged to verify that the average response time falls within the threshold limit based on various application cases. One of the canonical Distance Bounding Protocols, based on a shared symmetric key, was proposed by Hancke and Kuhn~\cite{hancke-kuhn-distance-bounding-2005} (see \autoref{fig:distance-bound-hanckle-kuhn}). The protocol begins with a session setup phase where the verifier and prover, which share a secret key $k$, exchange nonces $n_V$ and $n_P$ from a nonce space $\mathcal{N}$. The subsequent phase is the rapid exchange of bits for $N$ rounds which are timed by the verifier. The responses consist of the bits generated by a pseudo-random function outputting $2N$ bits using the nonces and the shared key to authenticate the prover. Let $c$ be the speed of light, $t_\mathsf{max}$ the time to process and release the response for a given challenge at the prover, and $d_\mathsf{max}$ the maximum approved distance between the prover and verifier for a specific application. The calculated round-trip response time at the verifier must be less than the approved threshold time: $t_\mathsf{thr} = 2*\frac{d_\mathsf{max}}{c} + t_{max}$. If all the responses are received within the threshold time and match with the local PRF calculation, the verifier approves the close proximity of the prover.

\section{Motivation and Problem Statement}

\para{User/Device Roles in Unlinkability} Throughout the paper, we will be using some of the key terms to describe our research work.  In this paper, the attacker's objective is to link two seemingly unrelated actions of the same user. The following are the user/device roles: A \textit{target device} belongs to the user whose privacy the adversary aims to violate. The owner of the target device will be referred to as the \textit{target user}. To break unlinkability, the adversary attempts to link the user of an arbitrary device, referred to as a \textit{prospective device}, to a user of the target devices. This set of target devices will be referred to as the \textit{target devices}.  The owner of a prospective device will be referred to as \textit{prospective user}. An adversary will test numerous prospective devices to find one that connects to the target device, thereby identifying the target user. \looseness=-1

\para{Motivating Example} We demonstrate two wireless protocol case scenarios with identical protocol designs, yet having different privacy impacts. Consider a database of passports that is queried at international airports. This database contains passport details of all countries (worldwide scope). On the other hand, consider two household Bluetooth devices: a cellphone and a speaker. Given their frequent connection, the cell phone has an allowlist (PNL) containing an entry for the speaker to facilitate faster reconnection. Unlike the worldwide scope of passport details, the scope of the Bluetooth allowlist database is limited to a household.
Both cases employ the Basic Hash protocol~\cite{basic-hash-protocol-source-2010}. In this protocol, the shared secret $k$ is hashed using a cryptographic hash function $\hash$ with a nonce $n$ as the tuple $\langle n,\hash(n, k)\rangle$. In,  Bluetooth's case the secret is sender device's $\mathsf{IRK}$, while the E-Passport reader uses pre-shared secrets over RFID communication.
In the attack, the adversary collects the message at time $t_1$ and attempts to replay it at time $t_2 > t_1$, either to the E-Passport reader at the airport or to an prospective user's cellphone in various locations.
At time $t_2$, a response of OK or ERROR from the passport reader only discloses that the target user information is present in a worldwide database. In contrast, for paired devices, an OK response provides more specific and sensitive information: the prospective user is part of the target user's household, violating location privacy by confirming the presence of the target user at the prospective device location. \looseness=-1

\para{Problem Statement}
Considering the above observation, this paper seeks to answer the following research questions (RQ):
\begin{itemize}[nosep]
\item [\textbf{RQ1.}] \textit{What causes PNL privacy attacks to work differently in differnt contexts?}
\item [\textbf{RQ2.}] \textit{How can these attacks be prevented formally and in practice?}
\end{itemize}

\para{Overview}
To address \textbf{RQ1}, we observe that the various allowlist network expose privacy-sensitive data differently.
In \S\ref{sec:new-privacy-vector-formalization}, we systematically describe the allowlist network, protocol checks, and adversary knowledge required for the PNL attack in terms of six necessary conditions. These conditions reveal unexplored privacy violations in protocols that use allowlists for peer agreements and authentication. To answer \textbf{RQ2}, we perform a formal security assessment of Bluetooth and \mbox{Wi-Fi} P2P (\S\ref{sec:case-studies}), leading to the identification of numerous old and new attacks.
Utilizing sufficient conditions from prior research and employing condition oblivious responses in conjunction with RR and DiB)properties, we counter relay/replay attacks to prevent user location tracking (\S\ref{sec:defense}). \looseness=-1

\section{Characterizing Association Inference Attack}
\label{sec:new-privacy-vector-formalization}

In this section, we characterize the attack into its building blocks. We discuss the shared data network, scope sensitivity, peristence and the protocol structure that enable the PNL attack. We also clarify the attacker resources and context accessible to the adversary.

\subsection{Infrastructure Requirements}

\label{sec:infrastructure-requirement}

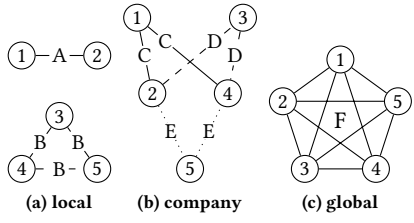
\begin{figure}[t] 
	\centering

	\subfloat[local]{
		\begin{tikzpicture}
		\tikzset{
    circled node/.style={circle,draw, inner sep=0, minimum size=1em,text height=.8em,text depth=.25em},
  }
		\node[circled node] (n1) at (0,0) {\small 1};
		\node[circled node] (n2) at (1,0) {\small 2};
		\draw (n1) --node[fill=white, circle, inner sep=0]{\small A} (n2);

		\node[circled node] (n3) at (0.5,-0.8) {\small 3};
		\node[circled node] (n4) at (0,-1.5) {\small 4};
		\node[circled node] (n5) at (1,-1.5) {\small 5};
		\draw[dashed] (n3) --node[fill=white, circle, inner sep=0]{\small B} (n4) -- node[fill=white, circle, inner sep=0]{\small B} (n5) -- node[fill=white, circle, inner sep=0]{\small B} (n3);
	\end{tikzpicture}
	}
	\subfloat[company]{
		\begin{tikzpicture}
		\tikzset{
    circled node/.style={circle,draw, inner sep=0, minimum size=1em,text height=.8em,text depth=.25em},
  }
		\node[circled node] (n2) at (0,0) {\small 2};
		\node[circled node] (n4) at (1,0) {\small 4};

		\node[circled node] (n1) at (-0.2,1) {\small 1};
		\node[circled node] (n3) at (1.2,1) {\small 3};
		\node[circled node] (n5) at (0.5,-1) {\small 5};
		\draw (n2) --node[fill=white, circle, inner sep=0]{\small C} (n1) -- node[fill=white,near start, circle, inner sep=0]{\small C} (n4);
		\draw[dashed] (n2) --node[fill=white,near end, circle, inner sep=0]{\small D} (n3) -- node[fill=white, circle, inner sep=0]{\small D} (n4);
		\draw[dotted] (n2) --node[fill=white, circle, inner sep=0]{\small E} (n5) -- node[fill=white, circle, inner sep=0]{\small E} (n4);
	\end{tikzpicture}
	}
	\subfloat[global]{
		\begin{tikzpicture}[scale=0.8]
		\tikzset{
    circled node/.style={circle,draw, inner sep=0, minimum size=1em,text height=.8em,text depth=.25em},
  }
		\node[circled node] (n1) at (90:1) {\small 1};
		\node[circled node] (n2) at (162:1) {\small 2};
		\node[circled node] (n3) at (234:1) {\small 3};
		\node[circled node] (n4) at (306:1) {\small 4};
		\node[circled node] (n5) at (18:1) {\small 5};

		\draw (n1) -- (n2) -- (n3) -- (n4) -- (n5) -- (n1) -- (n3) -- (n5) -- (n2) -- (n4) -- (n1);
		\node {F};
	\end{tikzpicture}
	}

	\caption{\small Different data share network. Nodes represent devices; edges indicate shared key data according to their letter.
	}
	\label{fig:data-share-network}
\end{figure}

In \autoref{fig:data-share-network} the sub-figures (a), (b), and (c) show an example of different shared data network among five devices $\{1...5\}$. In case (a), the devices $\{1,2\}$  and $\{3,4,5\}$ form separate connected components (local groups) and they share data $A$ and $B$, respectively. Examples of data in this network are allowlist in Bluetooth pairing and group-network- configuration in Wi-Fi P2P persistent group formation protocol. Here, the devices $\{3,4,5\}$ could be IoT devices of a house, with group network configuration files as data $B$ for fast reconnection. In case (b) A few devices $\{2, 4\}$ contain data for all the other devices $\{1,3,5\}$. Devices $\{2,4\}$ could be considered as global server hosting details for clients $\{1,3,5\}$. For example, $\{2,4\}$ could be DNS servers holding host domains to IP mappings data $C$, $D$, and $F$ for devices $\{1,3,5\}$, respectively. Another example could be global passport records servers $\{1,2,3\}$ holding bio information for users holding passports $\{1,3,5\}$. In case (c) a global data is shared among all devices. For example, data $F$ as root digital certificates of global certificate authorities for transport layer security protocol authentication shared among all Internet-connected devices $\{1...5\}$.

Our first observervation is that the network shown in \autoref{fig:data-share-network}(a) is vulnerable to PNL attacks while the other aren't. Assume a connected component in the network as the target group. In this case, if a PNL is shared within the target group, it allows the adversary to distinctly compare the communication pattern of a prospective device with the target group. More details will be explained in \S\ref{sec:attack-procedure-and-impact}.

Second, the scope of the shared data must be privacy sensitive. In \autoref{fig:data-share-network}(a), the target group could belong to various sizes of scope: a house, a university, small businesses such as restaurants and coffee shops, a city, countries, large corporate offices such as Google and Facebook, etc. The local scope is considered \emph{privacy sensitive} if the information about the association of an individual to the group can be reduced to a target user. The group of IoT devices in a house or in an office can be owned by an individual user, whereas a device group in a university cannot be necessarily owned by an individual. In this case, the former group is considered privacy-sensitive, and the latter is not. \looseness=-1

Third the shared data should not change across multiple sessions. A persistently shared data across multiple sessions is usually desired to authenticate the devices on a reconnection or to establish new secrets if needed. \looseness=-1

Lastly, the protocol for communication among the target group should emit responses dependent on the shared data. An example of such responses could be result of message integrity check (MIC) verification with a preshared key. Other example could be decryption and confirmation of a nonce sent recently in a protocol with using preshared key. The immediate response from the protocol agent executing the condition or any later responses at the agent can serve as the point for PNL attack. We will refer to such a condition as \textit{privacy-critical condition} in the rest of the paper. \looseness=-1

\subsection{Attacker's Resources} 
\label{sec:attackers-resources}

\para{Replay and Relay} The adversary requires knowledge of one or more static locations for the target device. For a replay attack, this location enables the adversary to collect messages for future replay. For a relay attack, the adversary utilizes the static location of the target device to establish a relay connection with an arbitrary prospective device located at any location. For instance, numerous users may be targeted by locating their houses or offices, as most IoT devices within a house are paired. \looseness=-1

\para{Observable Context} In terms of the context available to the adversary, note that the communication behavior between devices in the target devices deviates from that among unrelated devices. Specifically, responses at the privacy-critical conditions should be distinguishable when reconnecting to a target device versus an unrealed one. The differences may include response size, response time, message counts, encoding formats, or state transitions. \looseness=-1

\subsection{Threat Model and Scope}
\label{sec:threat-model}

\para{Threat Model} We adopt the standard Dolev-Yao adversary model~\cite{dolev-yao} where the adversary can observe, modify, block, and replay communication messages. The adversary is capable of relaying wireless packets by deploying multiple relay devices between two remote user devices. Additionally, the adversary can initiate a connection without involving device owner interaction. We exclude the leakage of protocol secrets (\eg preshared keys) or any compromise of the user devices in our model. In terms of timing side channels to breach privacy, we do not consider the adversary's ability to assess time differences to deduce unlinkability.

In the context of distance bounding protocols, various attacks have been classified based on compromised prover, compromised verifier, and secret key leakage. These include Mafia Fraud~\cite{distance-bounding-distance-fraud-1988}, Lone Distance Fraud~\cite{distance-bounding-distance-fraud-1988, distance-bounding-distance-hijacking-cremers-2012}, Distance Hijacking~\cite{distance-bounding-distance-hijacking-cremers-2012}, Terrorist Fraud~\cite{distance-bounding-distance-fraud-1988}, and Assisted Distance Fraud~\cite{distance-bounding-distance-hijacking-cremers-2012}. Since we do not consider compromised devices or secret key leakage, none of these attacks are relevant to our paper. Our case study uses the Bluetooth Specification V5.2 ~\cite{bt-spec-5.2}, \mbox{Wi-Fi} Protected Setup V2.0.8~\cite{wps-spec-2.0.8}, and \mbox{Wi-Fi} P2P Specification V1.9 ~\cite{wifi-direct-spec-1.9}. \looseness=-1

\para{Practicality of Threat Model}
Many wireless devices have designed reconnection procedures for frequently connecting devices, often based  on pre-shared keys, which facilitate automatic and quickconnections.
Relay attacks have been part of security research for the past 20 years, resulting in numerous practical attacks such as car theft featuring ``keyless'' entry~\cite{relay-carsteal-1-2011, relay-carsteal-2-2017} and remote pickpocketing through user's contactless payment cards~\cite{relay-pickpocket-2015} , as well as mafia fraud attacks over RFID~\cite{practical-rfid-mafia-attack-2009}. The relay capability is a standard threat model in various prior research on distance bounding protocols~\cite{distance-bounding-distance-fraud-1988, distance-bounding-distance-hijacking-cremers-2012, distance-bounding-mafia-attack-1988, distance-bounding-verify-wo-time-location-2018, distance-bounding-akiss-tool-2019, distance-bouding-with-rf-2010} and wireless relay/replay based attacks~\cite{relay-attacks-wifi-preauth-2023, relay-replay-attacks-short-range-system-2023, bat-attack-exclusive-use-2022, bt-pairing-5-attack-ndss-2023}. \looseness=-1

\para{Practicality of Secure Proximity Precision} Verifying physical proximity is critical in domains such as industrial RFID/NFC manufacturing and asset tracking, wireless payments, and autonomous vehicles. This necessity has prompted active research in this area, with distance bounding check implementations facing challenges such as precise time measurements and noise considerations. However, the need to prevent relay attacks has become vital in the era of wireless communication~\cite{dib-practical-1-2015, dib-practical-2-2012}. This is reflected in the availability of recent products such as the ATA5350 UWB (Ultra-Wideband) Transceiver~\cite{uwb-transceiver-implements-distance-bounding}, the NXP Mifare Plus EV2 Secure IC ~\cite{mifare-secure-ic-implements-distance-bounding}, and the EMV contactless payment standard~\cite{emv-payment-card-2009-implements-distance-bounding} that implement distance bounding to counter relay attacks. Similarly, although Bluetooth devices operate under low power, the need for distance bounding measurements has driven the development of the NXP KW36 Bluetooth Low Energy radio platform, which features distance bounding checks. The corresponding white paper~\cite{distance-bounding-for-bluetooth} demonstrates the feasibility and practicality of implementing distance bounding.   \looseness=-1

\subsection{Attack Procedure and Impact}\label{sec:attack-procedure-and-impact}

When the aforementioned infrastructure \S\ref{sec:infrastructure-requirement} and attacker's resources \S\ref{sec:attackers-resources} are present within the defined threat model, an adversary can exploit AInf to compromise a target user's unlinkability. An attacker can launch the AInf attack as follows: First, the attacker identifies a static location for a target user who owns target devices satifying infrastructure requirements \S\ref{sec:infrastructure-requirement}. Upon encountering a prospective device, the adversary is interested in determining whether the prospective device belongs to the target user.

For the replay attack, the adversary records the wireless communication within the target devices and later uses it to replay the messages to the prospective device. In a relay attack, the adversary establishes a relay connection between the target device and the prospective device. Based on the positive/negative conditional outcome, the adversary infers whether the prospective device is paired with a target device. If so, the target user is present at the location of the prospective device. Note that if a protocol contains multiple privacy-critical conditions exposing distinguishing behaviors, the adversary can use any of these conditions independently to carry out our AInf attack. For example, if for any privacy-critical condition, the connection proceeds as a reconnection procedure among target devices, it indicates that the prospective device is carried by the target user. On the other hand, for any progression of the connection that does not follow this pattern, the prospective device does not belong to the target user, and the adversary continues hunting for other prospective devices, repeating the same attack procedure.
The association of a prospective device with another target one allows the adversary to link the location and service interest of the target user, thereby undermining their unlinkability. For example, frequent visits to special medical facilities could reveal the physical conditions of the target user. \looseness=-1

For cases when the group of devices belongs a common theme ( such as a university, recreational club, religious organization, or specialized medical facility), an association inference of a prospective user to such a group can reveal information about their relationships, interests, or physical conditions.  However, this setting does not enable the tracking of an individual user's location. Instead, the privacy exposure for each individual user is $1/N$, where $N$ represents the number of target device owners. Therefore, as $N$ increases, the overall privacy impact will be reduced.

\section{Formalization}
In this section, we extend existing formalizations to capture our allowlist-based linkability attack.
We employ a similar process algebra as in \cite{privacy-3-conditions-tamarin-2020} to describe the possible protocol operations.
To emphasize the differences introduced by the use of allowlists, we present a minimal extension of their Simple Hash protocol.
Instead of a single global reader, we split the reader into two, each with its own database to illustrate the issue arising with allowlists.

\begin{figure}
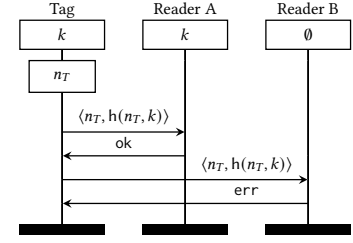

\begin{center}
\vspace{-8mm}
\scalebox{0.7}{
\begin{msc}[
instance distance=0.7cm, draw frame=none, msc keyword={},title top distance=0pt,title distance=0pt,
	first level height=2mm, level height=4.5mm
	]{}
\declinst{t}{Tag}{$k$}
\declinst{ra}{Reader A}{$k$}
\declinst{rb}{Reader B}{$\emptyset$}
\action{$n_T$}{t}
\nextlevel[3]
\mess{$\langle n_T,\hash(n_T,k)\rangle $}{t}{ra}
\nextlevel

\mess{\texttt{ok}}{ra}{t}
\nextlevel[1]
\mess{$\langle n_T,\hash(n_T,k)\rangle $}{t}[.75]{rb}
\nextlevel

\mess{\texttt{err}}{rb}[.25]{t}
\end{msc}}
\vspace{-6mm}
\end{center}
\caption{Example of attack with multiple readers}
\label{fig:multi-reader-allowlist}
\end{figure}

In this example, a \emph{tag} has a key $k$ and uses it to authenticate to a reader by sampling a nonce $n_T$ and sending the tuple $\langle n_T,\hash(n_T,k)\rangle$ to the reader. If the reader has $k$ in its allowlist, and the hash matches, it replies \texttt{ok}, otherwise \texttt{err}.
To demonstrate the effect of allowlists, we model two separate readers with their own databases acting as allowlists, i.e. sets of keys as shown in \autoref{fig:multi-reader-allowlist}.

In this minimal example, we simplify the complexity of a system like Bluetooth pairing by only authenticating the tag and omitting the reverse authentication. Instead of multiple pairings between devices, there are only two groups of readers in this model.

\subsection{Syntax}

We model these processes using process algebra with the following syntax.
Let $\mathcal{N}$ be a set of names, $\mathcal{X}$ variables and $\mathcal{C}$ channels. With these, $\mathcal{T}(\Sigma,V)$ creates terms using functions from the signature $\Sigma$ and data $V$. The full syntax is presented in \autoref{tab:syntax}.

The main change is the parameter $r$ to $\pins$ and $\plook$ which specifies the reader to act on. The reader identities $\mathsf{A,B}$ are public constants.

\begin{table}
\caption{Syntax of processes, $r\in\{\mathsf{A,B}\}$}
\label{tab:syntax}
\begin{tabular}{lll}
 $P,Q:=0$ & null process & \\
 $| \pn n.P$ & name restriction & $n\in\mathcal{N}$\\
 $|\pin(c,x).P$ & input & $c\in\mathcal{C}, x\in\mathcal{X}$\\
 $|\pout(c,u).P$ & output & $c\in\mathcal{C}, u\in\mathcal{T}(\Sigma_c,\mathcal{N}\cup\mathcal{X})$\\
 $|\pins_r(v)$ & insert to database $r$ & $v\in\mathcal{T}(\Sigma_c,\mathcal{N}\cup\mathcal{X})$\\
 $|\plook_r\ k $ &pairing test & $k\in\mathcal{N}$\\
 \quad such that $\bar x=\bar t$ & & $\bar x\in\mathcal{X}^{l}$ \\
 \quad in $P \pelse Q$& & $\bar t\in\mathcal{T}(\Sigma,\mathcal{N}\cup\mathcal{X})^l$\\
 $|(P|Q)$ & parallel composition\\
 $|!P$ & replication\\
 $|P;Q$ & sequence\\
 $|\repet P$ & repetition\\
\end{tabular}
\end{table}

\subsection{Semantics}

To maintain a record of allowlists and the adversary's knowledge, we define our system configurations $K$ as follows: $(\procs;\phi; (A_A,A_B))$
where $\procs$ represents the multiset of processes without any null processes.
$\phi=\{w_1\mapsto u_1,\dots,w_n\mapsto u_n\}$ is a substitution of handles $w_i$ to messages $u_i$ modelling the information available to the adversary, referred to as the \emph{frame}. And $(A_A,A_B)$ denote the two allowlists of the readers, for example, $A_A=\{k_1,k_2\}$. We write the combined tuple as $A:=(A_A,A_B)$
 With this configuration, we define our process semantics as the configuration transitions shown in \autoref{tab:semantics}.

 \begin{table}
  \caption{Semantics for processes}
\label{tab:semantics}
 \begin{tabular}{ll}
 IN & $(\pin(c,x).P\cup\procs;\phi;A) $ \\
 & $\xrightarrow{\pin(c,R)} (P\{x\mapsto u \}\cup\procs; \phi;A)$ \\
 &($R$ a recipe for the computable message $u$)\\
 OUT & $(\pout(c,u).P\cup\procs;\phi;A) $\\
 &$\xrightarrow{\pout(c,w)} (P\cup\procs; \phi\cup\{w\mapsto u\}; A)$\\
 NEW-N & $(\pn n.P\cup\procs; \phi; A) \xrightarrow{\tau} (P\cup\procs;\phi;A)$\\
 INSERT & $(\pins_r(v)\cup\procs; \phi; (A_r,\cdot)) $ \\
 & $\xrightarrow{\tau} (P\cup\procs;\phi;(A_r\cup\{v\},\cdot))$\\
 ALLOWED & $(\plook_r\ k \text{ such that }\bar x=\bar t \pin P \pelse Q\cup\procs;\phi;A)   $ \\
 -THEN& $ \xrightarrow{\tau_\pthen} (P\{k\mapsto k'\bar x \mapsto \bar w \}\cup\procs;\phi;A) $ \\
 & when $k'\in A_r$ exists such that $\bar w=\bar t\{k\mapsto k'\}\Downarrow$\\
ALLOWED & $(\plook_r\ k \text{ such that }\bar x=\bar t \pin P \pelse Q\cup\procs;\phi;A)   $ \\
 -ELSE& $ \xrightarrow{\tau_\pelse} (Q\cup\procs;\phi;A) $ \\
 & when $\forall k'\in A_r$, it holds that $\bar t\{k\mapsto k'\}\not\Downarrow$\\
 PARALLEL & $(\{P_1|P_2\}\cup\procs;\phi;A) \xrightarrow{\tau}(\{P_1,P_2\}\cup\procs;\phi;A)$\\
 REPLICATE & $(!P\cup\procs;\phi;A) \xrightarrow{\tau}(P\cup !P\cup\procs;\phi;A)$\\
 REPEAT & $(\repet P\cup\procs;\phi;A) \xrightarrow{\tau}(\{P; \repet P\}\cup\procs;\phi;A)$\\
 SEQ & $(P\cup\procs;\phi;A) \xrightarrow{\alpha}(P'\cup\procs;\phi';A')$ \\
 &for any sequence simplification rule \\
 ABORT & $((P;Q)\cup\procs;\phi;A) \xrightarrow{\tau_\mathsf{abort}}(Q\cup\procs;\phi;A)$ \\
 \end{tabular}
 \end{table}

\subsection{Protocols}

Using the algebra defined above, we model the real world execution as follows. Given a tag and receiver processes parametrized by $r$ as
\begin{align*}
T := &\pout(c_T,\langle n_T,h(k,n_T)\rangle)\\
R[r] :=&\pin(c_T,z).\plook_r k \st \bar x = \mathsf{eq}(\snd(z), h(k,\fst(z))) \\
&\text{ in } \pout(c_r,\mathtt{ok}) \pelse \pout(c_r,\mathtt{err})
\end{align*}
One tag and two receivers then form a real-world protocol
\begin{align*}
  \mathcal{M}_\Pi := &!(\pn k.\pin(c_X,r).\pins_r(k).\repet \pn n_T T ) |(! R[\mathsf{A}]) | (! R[\mathsf{B}])
\end{align*}
This allows for the creation of as many tags as needed, as well as multiple sessions of both reader types. Upon tag creation, it is paired with a reader chosen by the adversary, supplied through the channel $c_X$. Once set up, the tag repeats many sessions, each with a fresh nonce $n_T$.

Our next objective is to establish a protocol that only has a global reader.
To eliminate all effects from the allowlists, we define an ``unlinkable by definition'' protocol by creating a single receiver (wlog A) and disregarding the input on channel $c_X$:
\begin{align*}
  \mathcal{S}_\Pi := &!(\pn k.\pin(c_X,r).\pins_A(k).\repet \pn n_T T )|(! R[A])
\end{align*}
We argue that if an adversary can distinguish between the processes $\mathcal{M}_\Pi$ and $\mathcal{S}_\Pi$, it can link tags using the allowlists.

Using the initial example, we observe one possible execution trace:
$\tau_\mathsf{repl}.\tau_\mathsf{new}.\pin(c_X,\mathsf{A}).\tau_\mathsf{insertA}.\tau_\mathsf{repeat}.\tau_\mathsf{new}.\pout(c_T,w_0).\tau_\mathsf{seq}.\tau_\mathsf{repl}\\.\pin(c_T,z).\tau_\mathsf{then}.\pout(c_A,w_1)$
Continuing this trace with a subsequent interaction of tag $k'$ with $R[B]$ leads to the following continuation of the trace:
$.\tau_\mathsf{repeat}.\tau_\mathsf{new}.\pout(c_T,w_2).\tau_\mathsf{seq}.\tau_\mathsf{repl}.\pin(c_T,z).\tau_\mathsf{else}.\pout(c_B,w_3)$
The resulting frame is
\[\{w_0\to \langle n'_T,h(k',n'_T)\rangle, w_1\to \mathtt{ok},w_2\to \langle n''_T,h(k',n''_T)\rangle, w_3\to \mathtt{err}\}
\]

\begin{figure}
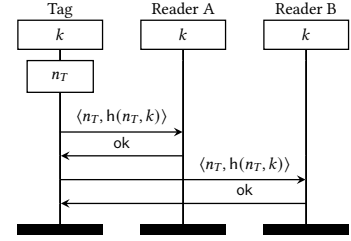

\begin{center}
\vspace{-8mm}
\scalebox{0.7}{
\begin{msc}[instance distance=0.7cm, draw frame=none, msc keyword={},title top distance=0pt,title distance=0pt,
	first level height=2mm, level height=4.5mm]{}
\declinst{t}{Tag}{$k$}
\declinst{ra}{Reader A}{$k$}
\declinst{rb}{Reader B}{$k$}
\action{$n_T$}{t}
\nextlevel[3]
\mess{$\langle n_T,\hash(n_T,k)\rangle $}{t}{ra}
\nextlevel

\mess{\texttt{ok}}{ra}{t}
\nextlevel[1]
\mess{$\langle n_T,\hash(n_T,k)\rangle $}{t}[.75]{rb}
\nextlevel

\mess{\texttt{ok}}{rb}[.25]{t}
\end{msc}}
\vspace{-6mm}
\end{center}
\caption{Unified reader database example}
\label{fig:unifiedDB}
\end{figure}

In the unlinkable scenario with a single reader, or two readers having the same allowlist (as shown in \autoref{fig:unifiedDB}), there will be no \textsf{err} return, and the frame for the same observable communication trace in $\mathcal{S}_\Pi$ is:
\[
\{w_0\to \langle n'_T,h(k',n'_T)\rangle, w_1\to \mathtt{ok},
w_2\to \langle n''_T,h(k',n''_T)\rangle, w_3\to \mathtt{ok}\}
\]

Since the outputs of cryptographic hash functions and nonces are indistinguishable, it is true that the parts of the resulting frames are statically equivalent:
\begin{align*}
\{ w_0\to \langle n'_T,h(k',n'_T)\rangle,w_2\to \langle n''_T,h(k',n''_T)\rangle\} \\
\sim \{ w_0\to \langle n'_T,h(k',n'_T)\rangle,w_2\to \langle n''_T,h(k',n''_T)\rangle\}
\end{align*}
However, the different remaining reply messages break the static equivalence:
\[\{w_1\to \mathtt{ok}, w_3\to \mathtt{err}\}\not\sim\{w_1\to \mathtt{ok}, w_3\to \mathtt{ok}\}\]
An adversary can exploit this discrepancy to distinguish the setting they are in and use it to link tag interactions.

\section{Proposed Mitigation}
\label{sec:defense}

To protect against AInf attacks, we employ a two-staged strategy combining: a) Prevention and Trapping, b) Detection. We describe these mechanisms in reverse order.

\subsection{Detection}

The purpose of this stage is to deterministically identify replay and relay attacks, enabling safe discconnection or continuation of the connection.

\para{Replay} A common method to prevent replay attacks is the use of counters, requiring synchronization between communicating devices to detect and discard replayed messages. However, this approach is susceptible to ND privacy violations~\cite{privacy-3-conditions-tamarin-2020}. Therefore, every reconnection establishes a fresh session key. Subsequent messages encrypted and integrity protected by this key cannot be replayed without being detected. On a reconnection with an allowlist, a mutually agreed fresh session key is derived by the end of the 3-round Authenticated Key Exchange (3AKE)~\cite{Boyed-Protocol-Classfication-IEEE-1995,Janson-and-Tsudik-3AKE-Serverless-cc-1995}. Our proposed protocol for A as initiator and B as responder combines preferred device verification within the 3AKE. Although the session key is only agreed upon by the end of the third round message, the key is generated at B after receiving the first message and after receiving a message at A in the second round. Therefore, all messages at the second round and onwards are sent ephemerally encrypted and integrity-protected. After mutual authentication with the 3AKE, a replay is detected deterministically.

\para{Relay} A common remedy for relay attacks is the use of GPS to determine the proximity of peer devices. However, the insufficient accuracy of GPS for this purpose~\cite{GPS-inaccurate-2015, GPS-inaccurate-II-2009} leads us to adopt distance bounding protocols~\cite{distance-bounding-herirachy-2018}. This protocol measures the distances of peer devices based on round-trip response delay to perform distance bounding (DiB). One caveat of DiB is that it requires authentication of the peer holding the same shared key (preferred device verification).
Therefore, we perform DiB immediately after the 3AKE as described above. In this way, the handshake strategically combines the setup phase for DiB~\cite{hancke-kuhn-distance-bounding-2005}, preferred device verification, and fresh session key establishment all together.
In this way, by the end of the DiB, agent A deterministically detects replay and relay attempts.

\subsection{Prevention and Trapping}

The goal of this stage is to prevent the exposure of any differentiable context during the 3AKE and to trap the adversary until the detection phase.

Recall that our allowlist privacy violation allows an adversary to infer associations by observing positive or negative conditional outcomes.
In our design, regardless of conditions' outcomes, responses are sent. In the positive outcome, the messages are encrypted and integrity protected as per the specification. In the case of a negative condition outcome, responses are indistinguishable, random values of the same size and count as a positive conditional outcome to avoid exposing outcomes via other side channels employing oblivious responses.

Further, we require the protocol to send messages that are encrypted by a fresh key and integrity-protected immediately after the session key generation. These messages are unique for the current session, hence cannot be replayed (unique response). Further, such messages cannot be predicted by the adversary (unpredictable response). Overall, we use fresh key bound messages to prevent exposing any relationship among messages, satisfying FO. Combining oblivious, unique, and unpredictable responses during 3AKE, prevents exposing differentiable context even if the adversary replays any message during the handshake. Therefore, the adversary is forced to continue to the later phases to find differentiable context, where, due to the mutually authenticated session key, any prior or recent replay is detected deterministically. At the same time, DiB detects the relay.

\looseness=-1

As such, our approach is to employ prevention and trapping until after 3AKE and DiB are completed, whereafter both replay and relay attacks can be detected within the protocol, enabling devices to disregard further messages from adversaries without compromising privacy. The salient feature of our design is that it keeps the faster reconnection requirements intact and provides a solution without needing synchronisation.

\section{Tamarin Model Design}
\label{tamarin-model-design}

As explained in \S\ref{sec:infrastructure-requirement}, modeling the formation of paired groups, un-paired groups; identities of users and devices; and shared data is critical to AInf. Precise protocol behavior and privacy properties are also essential. To achieve this, we use Tamarin \emph{Fact} and its associated Tamarin \emph{Rule} instantiation logic~\cite{tamarin-manual} to establish a hierarchical relationship among users, devices, protocol sessions, and the association of shared data/keys to the corresponding entities. Our model allows for an unbounded number of users and their respective devices. These devices interact with each other, executing an unbounded number of sessions based on the shared secrets stored as allowlists. The shared data is generated for each pair of devices associated with a user. This data is tied to the respective user and device identities. The sessions maintain the associated user and device identity terms through memory \emph{Facts}. These sessions use the identity terms to validate messages such as message integrity checks and decryptions with shared keys. Further, Tamarin \emph{Facts} model a database accessible to associated users and devices. In this way, the data is only accessed  by the devices within an allowlist group. Pattern matching~\cite{tamarin-manual}  is used with sessions to allow data access based on group hierarchy and device access control. For example, consider devices 1,2 of user $\mathsf{Alice}$ and devices 3,4 of user $\mathsf{Bob}$. The Tamarin persistent \emph{Fact}, with shared data $ \mathsf{LTK}_A $, is generated as $!\mathsf{DB}(\mathsf{Alice}, 1, \mathsf{LTK}_A)$ and $!\mathsf{DB}(\mathsf{Alice}, 2, \mathsf{LTK}_A)$. These \emph{Facts} represent the shared data $\mathsf{LTK}_A$ accessible only by devices ${1,2}$ of user $\mathsf{Alice}$. Similarly, for $\mathsf{Bob}$ the database \emph{Facts} $!\mathsf{DB}(\mathsf{Bob}, 3, \mathsf{LTK}_B)$ and $!\mathsf{DB}(\mathsf{Bob}, 4, \mathsf{LTK}_B)$ act as local storage for user $\mathsf{Bob}$. Devices of $\mathsf{Alice}$ cannot access shared secrets $\mathsf{LTK}_B$ of $\mathsf{Bob}$ by means of pattern matching.

By establishing the correct entity relationships and access control, as described above, allows encoding of required privacy properties. To prove WA, ND, and FO properties, we follow the symbolic formulation of ~\cite{privacy-2-condition-proverif-2016, privacy-3-conditions-tamarin-2020}. However, in order to secure privacy in paired communication, we formulate a trace-based property (termed as \emph{Lemma} in Tamarin): oblivious and unpredictable responses. Additionally, we adapt the FO modeling.

\begin{lemma}[Unique Unpredicable Responses]
\label{lemma:unun-responses}
Given labeled traces where every agent's response message $r_i$ from session $s_i$ to the network is labeled $\mathsf{Res}(s_i,r_i)$ at timestep $t_i$ and the set of attacker derivable terms is $\mathsf{KU}$

A protocol has Unique and Unpredictable Responses if responses are \emph{unique}: none of the privacy-critical responses are repeated:

$\forall s_1,s_2,r,t_1,t_2: \mathsf{Res}(s_1,r)@t_1 \land \mathsf{Res}(s_2,r)@t_2 \Rightarrow t_1=t_2$

and \emph{unpredictable}: none of the privacy-critical responses can be constructed by adversaries before their release:

$\land \forall s,r,t_i,t_j: \mathsf{Res}(s,r)@t_i \Rightarrow \not\exists (\mathsf{KU}(r)@t_j \land t_j < t_i)$
\end{lemma}

The DiB properties are obtained from prior research~\cite{lowe-auth-heirarchy, distance-bounding-verify-wo-time-location-2018}. Overall, to reason about privacy in our Tamarin model, we add the following security properties and revisions:  \looseness=-1

\para{Distance Bounding} Although DiB properties involve location and time measurements, Mauw et al.~\cite{distance-bounding-verify-wo-time-location-2018}  proposes a symbolic formulation for the DiB property without using location and time. In particular, it proves that as long as the response of a proximity check challenge cannot be sent before receiving the challenge itself, the protocol preserves a secure distance bounding check. Intuitively, an adversary can bypass the proximity check only if they can obtain protocol secrets to derive the DiB responses. Therefore we encode the DiB Lemma as follows:

\begin{lemma}[Distance Bounding]
Given annotated traces using the label $\mathsf{VS}(s,c)$ as the start of a verifier with session $s$ and challenge $c$, which waits for a corresponding response $r$ labeled $\mathsf{VE}(s,r)$. The honest prover from the same session $s$ emits a response labeled $\mathsf{P}(s,r)$. A protocol is Distance Bounding if $\forall s,c,r,t_s,t_e,t_p,t_k,s',t_{e'}$ it holds that for every verifier's released challenge $c$ and received a response $r$:

$\mathsf{VS}(s,c)@t_s \land \mathsf{VE}(s,r)@t_e \Rightarrow$

$\exists \mathsf{P}(s,r)@t_p$ \emph{the prover is alive with the corresponding response at time $t_p$}

$\land t_s < t_p < t_e$ \emph{the prover alive time $t_p$ is between released $t_s$ and received $t_e$ times}

$\land \not\exists (\mathsf{KU}(r)@t_k \land t_k<t_p)$ \emph{and the adversary cannot construct a correct response $r$ before the prover at $t_k$}

$\land \not\exists (\mathsf{VE}(s',r)@t_{e'} \land t_{e'}\not= t_e)$ \emph{and each response is uniquely verified only once without replay ($r$ is invalid in every other session $s'$)}
\end{lemma}

\looseness=-1

\para{Revised Frame Opacity Modeling}  Prior research~\cite{privacy-2-condition-proverif-2016, privacy-3-conditions-tamarin-2020} emphasizes the FO property to prevent an attack that exploits the relationship between messages. However, it does not replace public constants with idealized random values for diff-equivalence checking. In the context of communication among paired devices, any plaintext response from protocol agents provides a clear signal for a successful or unsuccessful condition. Therefore, our FO forces all responses from the protocol to be indistinguishable from nonces.  Finally, some protocols respond to the failure of verification checks by silently discarding the received message and stopping further communication. Since a silent discard response does not change among different protocol sessions, it forms an equality relationship of reactions in the protocols. Therefore, we can derive a FO privacy violation by modeling the silent discard behaviour as outputting a message of the public constant \texttt{SILENT\_DISCARD}.

It is worth noting that our oblivious and unpredictable response \autoref{lemma:unun-responses} ensures a similar goal as the FO property -- no relationship among messages -- but in the form of a trace property.  Since the FO property is proved in a separate model with omitted conditions (\S\ref{sec:threat-model}), we prove our oblivious and unpredictable response lemma as a trace property with conditions intact. \autoref{lemma:unun-responses} compliant responses are required from the first privacy-critical condition until the checkpoint where replay and relay can be detected. In order to encode this property, we designate the Tamarin \emph{Action Fact} $\mathsf{PCCResponse(session, response)}$ for all the individual outbound messages from both agents of the protocol. The response should be oblivious to preserve the FO, and it should be unpredictable to prohibit the adversary from gaining private information.\looseness=-1

\para{Tamarin Model Assumptions} For our Tamarin model, we simplify the protocol design by omitting the exchange of public and privacy-insensitive messages, including channel selection, supported bit rates, vendor, and manufacturing details, etc. The DiB check procedures in distance bounding protocols often require multiple rounds of challenges and response exchanges to improve statistics. However, following the formal verification of distance bounding by ~\cite{distance-bounding-verify-wo-time-location-2018}, we simplify it to only one round of challenge and keyed hash response. As described by Baelde et al., modeling FO requires removing privacy critical condition checks in the modeled protocol agents, thereby proving FO for a more generic protocol~\cite{privacy-3-conditions-tamarin-2020}. In this way, FO is verified for a superset of desired traces, implying verification on the desired subsets without loss of generality. Note that this applies only to the FO property. \looseness=-1

\section{Case Studies}
\label{sec:case-studies}

In this section, we explain our analysis of two real-world wireless communication protocols and describe the identified vulnerabilities utilizing formal modeling.

\subsection{Bluetooth}

\newcommand{\offsetangle}{0}
\tikzset{
  msc sectionmaker/.style={
    solid,
    inner sep=0.5,
    fill=blue,
    text=white,
    minimum size=.05cm,
    scale=0.8,
    shape=semicircle,
    shape border rotate=0
  }
}

As depicted in \autoref{fig:bleutooth-advertisement-n-SKD}, the Bluetooth reconnection procedure starts with the advertisement of Random Private Addresses (RPA). Party activations \encircleB{1} (between \tikz{\node[msc sectionmaker,shape border rotate=0]{1};} and  \tikz{\node[msc sectionmaker,shape border rotate=180]{1};}) and \encircleB{2} indicate both $A$ and $B$ generating their respective RPA and verifying the received RPA in \encircleB{2} and \encircleB{3} respectively. If the \textsf{IRK} does not match any entry in the device's allowlist, the receiving device silently discards the received message and stops responding further~\cite{5.2,4.2-general}.

After successfully resolving the \textsf{IRK}, the session key derivation phase begins. Since this is a reconnection procedure, both devices share a \textsf{LTK} from the pairing operation (\S\ref{sec:background-related}). Both devices start by sharing nonces which are known as Session Key Diversifiers ($\mathsf{skd}$). We omit the details of shared nonce IV, because the symbolic randomness of $\mathsf{skd}$ in Tamarin subsumes the randomness of IV as well. Note that when $B$ responds with its nonce $\mathsf{skd}_B$, it also informs $A$ to start the encryption by sending a plaintext \texttt{START\_ENC\_REQ} message~\cite{bt-spec-5.2}. Thereafter, both $A$ and $B$ encrypt a static, known value with the derived session key $\mathsf{sessK}$. These packets are denoted by \texttt{START\_ENC\_RSP}. After the successful verification of encrypted values on both ends, secure data exchange follows. In the case of a MIC failure at \encircleB{6} or \encircleB{7}, the devices silently discard the received message and transition to their standby mode ("\textit{...it shall immediately exit the Connection state, and shall transition to the Standby state}" \cite[page.~3033]{bt-spec-5.2}). The term MIC is used to distinguish from MAC, which could be confused with Media Access Control (MAC) addresses.

\newcommand{\skd}{\mathsf{skd}}
\newcommand{\sessK}{\mathsf{SessK}}
\newcommand{\irk}{\mathsf{irk}}
\newcommand{\LTK}{\mathsf{LTK}}
\newcommand{\DB}{\mathsf{DB}}
\newcommand{\MIC}{\mathsf{MIC}}
\begin{figure}[t]
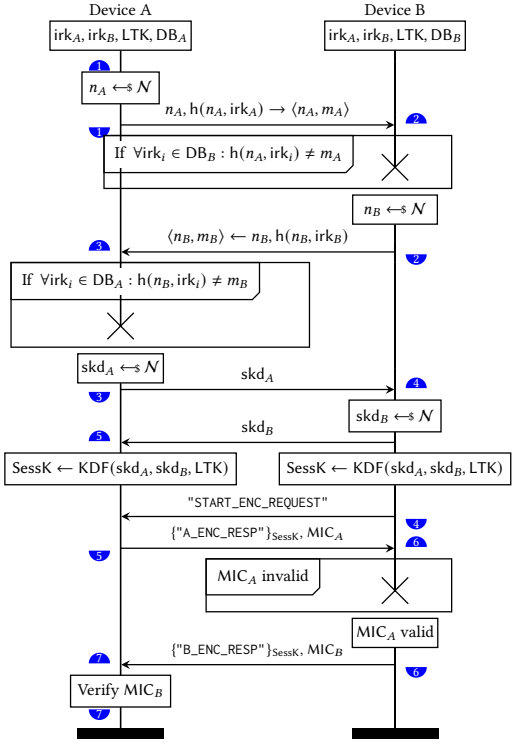
 
	\centering
\scalebox{0.7}{
	\begin{msc}[instance distance=2.5cm, draw frame=none, msc keyword={},title top distance=0pt,title distance=0pt,
	first level height=2mm, level height=2mm]{}
\declinst{a}{Device A}{$\irk_A,\irk_B,\LTK,\DB_A$}
\declinst{b}{Device B}{$\irk_A,\irk_B,\LTK,\DB_B$}
\msccomment[msccomment distance=-4mm,side=left]{1}{a}
\nextlevel
\action*{$n_A\sample\mathcal{N}$ }{a}
\nextlevel[4]
\msccomment[msccomment distance=4mm,side=left]{2}{b}%
\nextlevel[1]
\mess{$n_A,\hash(n_A,\irk_A)\to\langle n_A,m_A\rangle$}{a}{b}
\msccomment[msccomment distance=-4mm,side=right]{1}{a}
\nextlevel
\inlinestart[right inline overlap=10.7mm,left inline overlap=55mm]{condB}{\colorbox{white}{If $\ \forall \irk_i\in\DB_B : \hash(n_A,\irk_i) \neq m_A $}}{b}{b}
\nextlevel[3]
\stop{b}
\nextlevel[2]
\inlineend{condB}
\nextlevel[2]
\startinst{b}{}{$n_B\sample\mathcal{N}$}
\nextlevel[3]
\msccomment[msccomment distance=-4mm,side=left]{3}{a}
\nextlevel
\mess{$\langle n_B,m_B\rangle\gets n_B,\hash(n_B,\irk_B)$}{b}{a}
\msccomment[msccomment distance=4mm,side=right]{2}{b}
\nextlevel
\inlinestart[right inline overlap=35.7mm,left inline overlap=20.7mm]{condA}{\colorbox{white}{If $\ \forall \irk_i\in\DB_A : \hash(n_B,\irk_i) \neq m_B $}}{a}{a}
\nextlevel[6]
\stop{a}
\nextlevel[2]
\inlineend{condA}
\nextlevel[2]
\startinst{a}{}{$\skd_A\sample\mathcal{N}$}
\nextlevel[1]
\msccomment[msccomment distance=4mm,side=left]{4}{b}
\nextlevel
\mess{$\skd_A$}{a}{b}
\msccomment[msccomment distance=-4mm,side=right]{3}{a}
\nextlevel[1]
\action*{$\skd_B\sample\mathcal{N}$}{b}
\nextlevel[3]
\msccomment[msccomment distance=-4mm,side=left]{5}{a}
\nextlevel
\mess{$\skd_B$}{b}{a}
\nextlevel
\action*{$\sessK\gets\KDF(\skd_A,\skd_B,\LTK)$}{b}
\action*{$\sessK\gets\KDF(\skd_A,\skd_B,\LTK)$}{a}
\nextlevel[6]
\mess{\texttt{\small "START\_ENC\_REQUEST"}}{b}{a}
\msccomment[msccomment distance=4mm,side=right]{4}{b}
\nextlevel[2]
\msccomment[msccomment distance=4mm,side=left]{6}{b}
\nextlevel
\mess{\small \{\texttt{"A\_ENC\_RESP"}$\}_\sessK$, \textsf{MIC}$_A$}{a}{b}
\msccomment[msccomment distance=-4mm,side=right]{5}{a}
\nextlevel[1]
\inlinestart[right inline overlap=10.7mm,left inline overlap=35.7mm]{mic}{\colorbox{white}{\textsf{MIC}$_A$ invalid}}{b}{b}
\nextlevel[3]
\stop{b}
\nextlevel[2]
\inlineend{mic}
\nextlevel[2]
\startinst{b}{}{$\MIC_A$ valid}
\nextlevel[2]
\msccomment[msccomment distance=-4mm,side=left]{7}{a}
\nextlevel
\mess{\small \{\texttt{"B\_ENC\_RESP"}$\}_\sessK$, \textsf{MIC}$_B$}{b}{a}
\msccomment[msccomment distance=4mm,side=right]{6}{b}
\nextlevel[1]
\action*{Verify \textsf{MIC}$_B$}{a}
\nextlevel[3]
\msccomment[msccomment distance=-4mm,side=right]{7}{a}
\end{msc}}
\vspace{-5mm}

	\caption{Simplified Bluetooth advertisement and session key derivation protocol for reconnection procedure.}\label{fig:bleutooth-advertisement-n-SKD}
\end{figure}

\para{Requirements for the AInf Attack}
In the BAT attack, the shared data is the allowlist containing trusted device identities and keys: \textsf{IRK}s and \textsf{LTK}s. The allowlist is persistent across different Bluetooth communication sessions and local to only a few of the devices owned by a user forming a paired device group. The wireless communication medium and advertisement stages of Bluetooth communications allow an adversary to initiate a wireless connection without any user interaction. Finally, only the devices that share the same \textsf{IRK} and \textsf{LTK} can continue the connection. Other devices observe otherwise. Such deviating interactions of paired and known unpaired groups leak the information of devices belonging to the allowlist formed among IoT devices of the target user.

\para{The Discovered Privacy Vulnerabilities}
There are a few key checkpoints in the protocol that violate privacy. The advertisement messages at \encircleB{2} and \encircleB{3} are vulnerable to replay or relay of the RPAs as explained in the BAT attack~\cite{bat-attack-exclusive-use-2022}.  The Tamarin model reveals this by the failure of the WA and FO properties. FO is violated because of the silent discard, and WA is violated due to the possible replay of RPA messages. Similarly, at \encircleB{6} the \texttt{SILENT\_DISCARD} indicates different \textsf{LTK}s for each party. A relay adversary against privacy in the Bluetooth reconnection procedure uses the \texttt{START\_ENC\_REQ} response at \encircleB{4}, which indicates correctness of all earlier verification checks: Both RPA resolutions in the advertisement phase at \encircleB{2} and \encircleB{3}. Note that during the relay, the random and hashed responses at \encircleB{4} and \encircleB{5} are not an indication of the verification condition outcomes. In contrast, \texttt{START\_ENC\_REQ} is a plaintext message that clearly indicates the positive condition outcome of the prior resolution of RPAs. In the Tamarin model, this is captured through a FO counter example using the plaintext message. Vulnerabilities at \encircleB{4} and\encircleB{6} are newly discovered in this paper as shown in \autoref{tab:vulnerabilities-new-old}.

Note that all the checkpoints in the table independently contribute to our Ainf attack, and any one of them is sufficient. As mentioned earlier, some of these checkpoints were discovered in prior research; however, we provide new checkpoints deep into the sequences that expose our Ainf attack. Given the checkpoints’ independence, all must be mitigated to prevent the Ainf attack.

\subsection{Countermeasures for Bluetooth}
\label{sec:bt-proposed}

\begin{figure}[t]
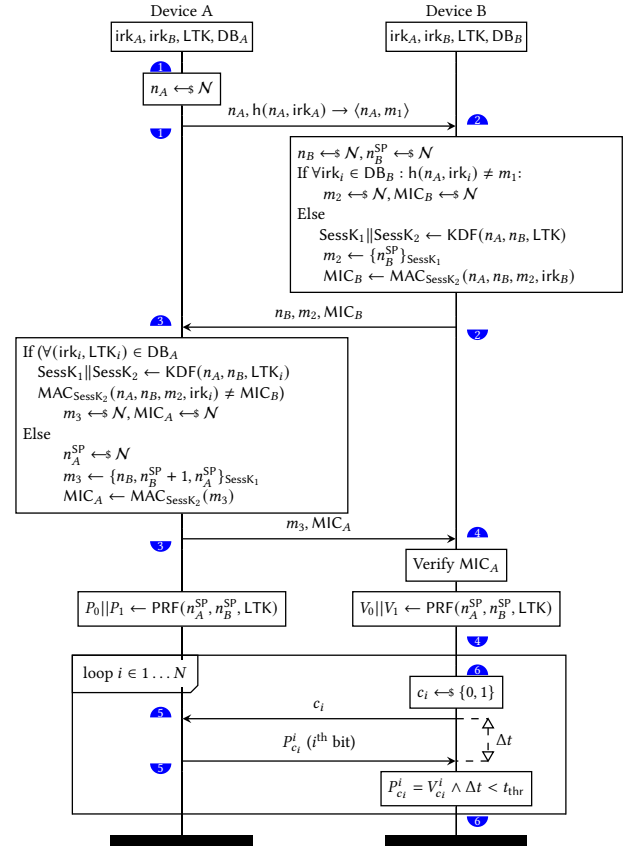

	\centering
	\vspace{-6mm}
\scalebox{0.7}{
	\begin{msc}[instance distance=2.5cm, draw frame=none, msc keyword={},title top distance=0pt,title distance=0pt,
	first level height=2mm, level height=2mm]{}
\declinst{a}{Device A}{$\irk_A,\irk_B,\LTK,\DB_A$}
\declinst{b}{Device B}{$\irk_A,\irk_B,\LTK,\DB_B$}
\msccomment[msccomment distance=-4mm,side=left]{1}{a}
\nextlevel
\action*{$n_A\sample\mathcal{N}$}{a}
\nextlevel[4]
\msccomment[msccomment distance=4mm,side=left]{2}{b}
\nextlevel
\mess{$n_A,\hash(n_A,\irk_A)\to\langle n_A,m_1\rangle$}{a}{b}
\msccomment[msccomment distance=-4mm,side=right]{1}{a}
\nextlevel[1]
\action*{\begin{minipage}{6cm}
	$n_B\sample\mathcal{N},n_B^\mathsf{SP}\sample\mathcal{N}$\\
	If $\forall \irk_i\in\DB_B : \hash(n_A,\irk_i) \neq m_1 $:\\
	\hspace*{3ex} $m_2 \sample \mathcal{N}, \MIC_B\sample\mathcal{N} $\\
	Else\\
    \hspace*{3ex}$\sessK_1\|\sessK_2 \gets\KDF(n_A,n_B,\LTK)$\\
	\hspace*{3ex} $m_2\gets\{n_B^\mathsf{SP}\}_{\sessK_1}$\\
	\hspace*{3ex} $\MIC_B\gets\mathsf{MAC}_{\sessK_2}(n_A,n_B,m_2,\irk_B)$
	\end{minipage}}{b}
\nextlevel[17]
\msccomment[msccomment distance=-4mm,side=left]{3}{a}
\nextlevel
\mess{$n_B,m_2,\MIC_B$}{b}{a}
\msccomment[msccomment distance=4mm,side=right]{2}{b}
\nextlevel[1]
\action*{\begin{minipage}{6cm}
    If ($\forall (\irk_i,\LTK_i)\in\DB_A$ \\
	\hspace*{2ex}$\sessK_1\|\sessK_2 \gets\KDF(n_A,n_B,\LTK_i)$\\
    \hspace*{2ex}$\mathsf{MAC}_{\sessK_2}(n_A,n_B,m_2,\irk_i) \neq \MIC_B$)\\
    \hspace*{5ex} $m_3 \sample \mathcal{N}, \MIC_A\sample\mathcal{N} $\\
	Else\\
	\hspace*{5ex} $n_A^\mathsf{SP} \sample \mathcal{N}$\\
	\hspace*{5ex} $m_3\gets\{n_B,n_B^\mathsf{SP}+1,n_A^\mathsf{SP}\}_{\sessK_1}$\\
	\hspace*{5ex} $\MIC_A\gets\mathsf{MAC}_{\sessK_2}(m_3)$
	\end{minipage}}{a}
	\nextlevel[18]
\msccomment[msccomment distance=4mm,side=left]{4}{b}
\nextlevel
\mess{$m_3,\MIC_A$}{a}{b}
\msccomment[msccomment distance=-4mm,side=right]{3}{a}
\nextlevel[1]
\action*{Verify $\MIC_A$}{b}
\nextlevel[4]
\action*{$P_0||P_1\gets\mathsf{PRF}(n_A^\mathsf{SP},n_B^\mathsf{SP},\LTK)$}{a}
\action*{$V_0||V_1\gets\mathsf{PRF}(n_A^\mathsf{SP},n_B^\mathsf{SP},\LTK)$}{b}
\nextlevel[4]
\msccomment[msccomment distance=4mm,side=right]{4}{b}
\nextlevel[2]
\inlinestart[right inline overlap=20.7mm,left inline overlap=20.7mm]{loop}{\colorbox{white}{loop $i\in 1\dots N$}}{a}{b}
\nextlevel
\msccomment[msccomment distance=4mm,side=left]{6}{b}
\nextlevel
\action*{$c_i\sample\{0,1\}$}{b}
\nextlevel[3]
\msccomment[msccomment distance=-4mm,side=left]{5}{a}
\nextlevel
\measure[side=right]{$\Delta t$}{b}{b}[4]
\mess{$c_i$}{b}{a}
\nextlevel[4]
\mess{$P_{c_i}^i$ ($i$\textsuperscript{th} bit)}{a}{b}
\msccomment[msccomment distance=-4mm,side=right]{5}{a}
\nextlevel
\action*{$P_{c_i}^{i}=V_{c_i}^{i}\land \Delta t<t_\mathsf{thr}$}{b}
\nextlevel[4]
\msccomment[msccomment distance=4mm,side=right]{6}{b}
\inlineend{loop}
\end{msc}
}
	\caption{Proposed Bluetooth advertisement and session key derivation protocol for the reconnection procedure.}\label{fig:bleutooth-advertisement-n-SKD-fix}
\end{figure}

We propose an updated protocol for Bluetooth advertisement and session key derivation (\autoref{fig:bleutooth-advertisement-n-SKD-fix}). We introduce a revised 3-way handshake, which ends with $A$ sending at \encircleB{3} and $B$ receiving at \encircleB{4}, followed by a DiB check. In the process, we add several critical modifications. Our modification begins at step \encircleB{2}, where device $B$ not only verifies $A$'s RPA. It derives a session key $\mathsf{sessK}$, combining its own nonce $n_B$ with $A$'s nonce $n_A$ and the shared \textsf{LTK} ($\mathsf{sessK} = \mathsf{KDF}(n_A, n_B, \textsf{LTK})$. Simultaneously, $B$ forgoes generating a hashed RPA and instead creates a MIC over $B$'s \textsf{IRK} using $\mathsf{sessK}$ as key. At step \encircleB{3}, device $A$ resolves the \textsf{IRK} by constructing a local MIC and comparing it with the received one. This method of resolving the \textsf{IRK} using a MIC facilitates a multi-step process: simultaneous resolution of \textsf{IRK} and \textsf{LTK}; derivation of the identical session key $\mathsf{sessK}$; and an integrity check for the proximity check nonce $n_\mathsf{B-SP}$. \looseness=-1

To impose oblivious and unpredictable responses, steps \encircleB{2} and \encircleB{3} are designed to transmit some data independent of whether the MIC verification succeeds or fails. In the failure cases, nonces with the same structure as successful response messages are returned. For example, at \encircleB{2}, the nonce MIC and $M2$ are crafted to match the size of the successful \textsf{IRK} resolution response messages. Uniformly random encrypted messages and MICs with the transient session key $\mathsf{sessK}$ facilitate oblivious responses across protocol sessions. Thus, even if $A$'s RPA is replayed at \encircleB{1}, adversaries will be thwarted from decoding the \textsf{IRK} resolution condition outcome. Moreover, as the responses at \encircleB{2} and \encircleB{3} are encrypted with the ephemeral session key, these messages are immune to replay. Ultimately, at step \encircleB{4}, $B$ can discern if any previous messages were replayed by verifying the authenticated agreement of the transient session key $\mathsf{sessK}$.
Finally, at \encircleB{4}, $B$ triggers a proximity check phase to validate $A$'s close proximity, enabling $B$ to detect both replay and relay attack vectors without exposing any associations.

\subsection{\mbox{Wi-Fi}}

\renewcommand{\offsetangle}{90}

\tikzset{
  msc sectionmaker/.style={
    solid,
    inner sep=0.5,
    fill=green,
    text=black,
    minimum size=.05cm,
    scale=0.8,
    shape=isosceles triangle,
  }
}

A simplified \mbox{Wi-Fi} P2P persistent reconnection procedure is shown in \autoref{fig:wifi-p2p-persistent-simplified}. Assume device $A$ acts as a \emph{supplicant} and device $B$ acts as an \emph{authenticator} (\S\ref{sec:background-related}). Similar to Bluetooth, there are two key phases: the advertisement and the session key derivation phase. In the former, $A$ seeks to join a previously connected group identified with $\mathsf{netID}$ at invocation \encircleW{1} (between \tikz{\node[msc sectionmaker,shape border rotate=-90]{1};} and \tikz{\node[msc sectionmaker,shape border rotate=90]{1};}).  If $B$ is part of this group,  it retrieves its stored configuration for the $\mathsf{netID}$ group at \encircleW{2}. If $A$ is also capable of acting as an authenticator, it proceeds to authenticate $B$.

\newcommand{\netID}{\mathsf{netID}}
\newcommand{\PSK}{\mathsf{PSK}}
\newcommand{\rc}{\mathsf{rc}}
\begin{figure}[t]
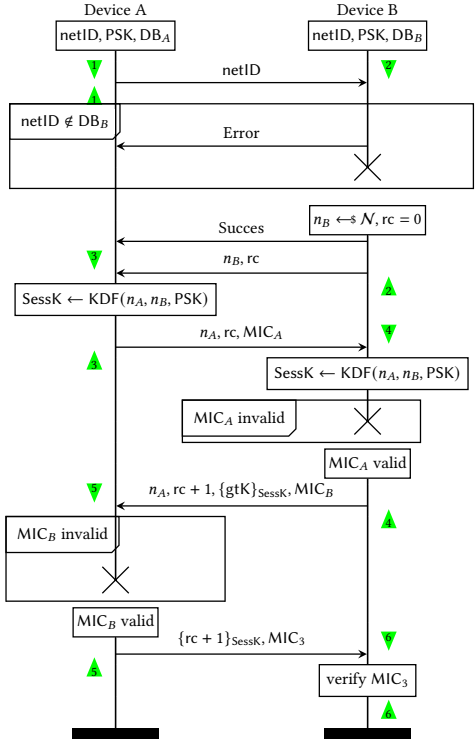
 
	\centering
	\scalebox{0.7}{
	\begin{msc}[instance distance=2.5cm, draw frame=none, msc keyword={},title top distance=0pt,title distance=0pt,
	first level height=2mm, level height=2mm]{}
\declinst{a}{Device A}{$\netID,\PSK, \DB_A$}
\declinst{b}{Device B}{$\netID,\PSK,\DB_B$}
\msccomment[msccomment distance=-4mm,side=left]{1}{a}%
\msccomment[msccomment distance=4mm,side=left]{2}{b}%
\nextlevel[2]
\mess{$\netID$}{a}{b}
\msccomment[msccomment distance=-4mm,side=right]{1}{a}%
\nextlevel[2]
\inlinestart[right inline overlap=20mm,left inline overlap=20mm]{condB}{\colorbox{white}{$\netID\not\in\DB_B$}}{a}{b}
\nextlevel[4]
\mess{Error}{b}{a}
\nextlevel[2]
\stop{b}
\nextlevel[2]
\inlineend{condB}
\nextlevel[3]
\startinst{b}{}{$n_B\sample\mathcal{N},\rc=0$}
\nextlevel[2]
\mess{Succes}{b}{a}
\nextlevel[1]
\msccomment[msccomment distance=-4mm,side=left]{3}{a}%
\nextlevel[2]
\mess{$n_B,\rc$}{b}{a}
\msccomment[msccomment distance=4mm,side=right]{2}{b}%
\nextlevel[1]
\action*{$\sessK\gets\KDF(n_A,n_B,\PSK)$}{a}
\nextlevel[4]
\msccomment[msccomment distance=4mm,side=left]{4}{b}%
\nextlevel[2]
\mess{$n_A,\rc,\MIC_A$}{a}{b}
\msccomment[msccomment distance=-4mm,side=right]{3}{a}%
\nextlevel[1]
\action*{$\sessK\gets\KDF(n_A,n_B,\PSK)$}{b}
\nextlevel[4]
\inlinestart[right inline overlap=10mm,left inline overlap=35mm]{condB2}{\colorbox{white}{$\MIC_A$ invalid}}{b}{b}
\nextlevel[2]
\stop{b}
\nextlevel[2]
\inlineend{condB2}
\nextlevel[2]
\startinst{b}{}{$\MIC_A$ valid}
\nextlevel[2]
\msccomment[msccomment distance=-4mm,side=left]{5}{a}%
\nextlevel[2]
\mess{$n_A,\rc+1,\{\mathsf{gtK}\}_\sessK,\MIC_B$}{b}{a}
\msccomment[msccomment distance=4mm,side=right]{4}{b}%
\nextlevel
\inlinestart[right inline overlap=20.7mm,left inline overlap=20.7mm]{condA}{\colorbox{white}{$\MIC_B$ invalid}}{a}{a}
\nextlevel[6]
\stop{a}
\nextlevel[2]
\inlineend{condA}
\nextlevel[2]
\startinst{a}{}{$\MIC_B$ valid}
\nextlevel[1]
\msccomment[msccomment distance=4mm,side=left]{6}{b}%
\nextlevel[2]
\mess{$\{\rc+1\}_\sessK, \MIC_3$}{a}{b}
\msccomment[msccomment distance=-4mm,side=right]{5}{a}%
\nextlevel[1]
\action*{verify $\MIC_3$}{b}
\nextlevel[3]
\msccomment[msccomment distance=4mm,side=right]{6}{b}%
\nextlevel
\end{msc}
}
	\vspace{-6mm}
	\caption{Simplified \mbox{Wi-Fi} P2P persistent group formation protocol flow.}\label{fig:wifi-p2p-persistent-simplified}
\end{figure} 

In the session key derivation phase, also known as the four-way handshake, $B$ generates and sends a random nonce $n_B$ and a replay counter $\mathsf{rc}$. $A$ receives $B$'s nonce $n_B$ and combines it with its own nonce $ n_A $ and the shared secret $\mathsf{PSK}$ to generate a session key $\mathsf{sessK} = \mathsf{KDF}(n_A, n_B, \mathsf{PSK})$ at \encircleW{3}. In response,  $A$ echoes back the replay counter $\mathsf{rc}$, nonce $n_A$, and MIC for the response. At \encircleW{4},   $B$ verifies the MIC and responds with the necessary group information (e.g., the Group Temporal Key $\mathsf{gtK}$). This key is used for sending secure and encrypted group messages in an established \mbox{Wi-Fi} P2P group.  The response at \encircleW{4}  includes the incremented value of $\mathsf{rc}$ and the nonce $n_B$.  $A$ verifies the MIC and acknowledges the message with an encryption of the incremented replay counter and the corresponding MIC. Note that the replay counter $\mathsf{rc}$ is increased by one after every pair of messages received and sent from either device. This is done to prevent replay of messages within the same session. \looseness=-1

\para{Requirements for the AInf Attack}
As per \S\ref{sec:infrastructure-requirement}, \mbox{Wi-Fi} P2P stores network IDs and the $\mathsf{PSK}$s as the allowlist in the local configuration file. This data persists across reconnecting procedures. Typically, IoT devices of a house form such persistent groups for faster reconnection of devices. Thus, our AInf attack on these network configurations exposes privacy-sensitive information about the ownership of the devices. Further, the wireless communication medium enables a device to initiate a reconnection without user involvement. The MIC verification conditions in the protocols serve as the privacy-critical conditions. Consequently, the \mbox{Wi-Fi} P2P persistent reconnection procedure is susceptible to our AInf attack.

\newcommand\featuretext[1]{%
  \llap{\vrule width.35pt height2pt depth2.5pt\kern1pt}%
  \rlap{\rotatebox{15}{\textbf{#1}}}%
}

\begin{table}[t]
\caption{Summary of the identified vulnerabilities. Acronyms: \textbf{BT:} Bluetooth, \textbf{WF:} \mbox{Wi-Fi}-P2P}
\label{tab:vulnerabilities-new-old}

\centering
\scriptsize
\setlength\tabcolsep{10pt}
 {%
\begin{tabular}{lccccc}
\featuretext{Protocols} & \featuretext{Checkpoint(s)} & \featuretext{Property  Failure} & \featuretext{Relay/Replay Attack}  & \featuretext{New?} & \featuretext{Known} \\ 
\midrule
BT & \encircleB{1}, \encircleB{2}, \encircleB{3}  & FO, WA  &  Both  & \xmark &  ~\cite{bat-attack-exclusive-use-2022}  \\
BT & \encircleB{4}  & FO  &  Relay  & \cmark & -- \\
BT & \encircleB{6}  & FO  &  Relay  & \cmark & -- \\
WF & \encircleW{1}, \encircleW{2}  & FO, WA  &  Both  & \xmark & ~\cite{ssid-deanonymize-I-2016, ssid-deanonymize-II-2015}  \\
WF & \encircleW{2}, \encircleW{3}, \encircleW{4}  & FO  &  Relay  & \cmark & -- \\
WF &  \encircleW{4} \encircleW{5}  & FO  &  Relay & \cmark & -- \\

\bottomrule[1.5pt]
\end{tabular}
}
\end{table}

\para{The Discovered Privacy Vulnerabilities}
The first privacy violation occurs when device $B$ confirms the presence of $\mathsf{netID}$ in its locally stored configuration. It responds with the plaintext status of the match. In case of a failure, the response is sent with the response code ``\textit{Fail; unknown P2P Group}''~\cite[page.~93]{wifi-direct-spec-1.9}. The plaintext responses expose an association between $A$ and the household's devices group of $B$. Therefore, the \mbox{Wi-Fi} P2P advertisement phase is vulnerable to both replay and relay attacks due to the failure of FO, RR, and WA. Note that \mbox{Wi-Fi}'s plaintext SSID broadcast privacy violations were already studied~\cite{ssid-deanonymize-I-2016, ssid-deanonymize-II-2015}.

Recall that in the relay attack, any positive or negative indication of a conditional outcome can be inferred through later messages in the protocol. Particularly, the sequence of the replay counter $\mathsf{rc}$, which is echoed back at \encircleW{3} to $A$, and again the incremented value $\mathsf{rc}+1$ at \encircleW{4} to $B$. Although successive $\mathsf{rc}$ values are designed to prevent message replay, their plaintext representations cause privacy violations. Moreover, the plaintext echo of the nonce $ n_A $ is returned at \encircleW{3} and \encircleW{4}. This behavior violates the FO privacy property. The sequential relationship between the replay counter values and the equality of the nonce $ n_A $ imply a successful pass of prior privacy-critical conditions. Such a violation is similar to the plaintext response at \encircleB{4} in the Bluetooth reconnection procedure. Finally, \texttt{SILENT\_DISCARD} at \encircleW{4} and \encircleW{5} of both devices~\cite[page.~2560-2564]{wps-spec-2.0.8} further violates the FO property. Note that vulnerabilities due to the sequential message relationship at \encircleW{2}, \encircleW{3}, \encircleW{4} and silent discard at \encircleW{4}, \encircleW{5} are newly discovered in this paper, as shown in \autoref{tab:vulnerabilities-new-old}. \looseness=-1

\subsection{Countermeasures for \mbox{Wi-Fi}}
\label{sec:wifi-p2p-proposed}

\begin{figure}[t]
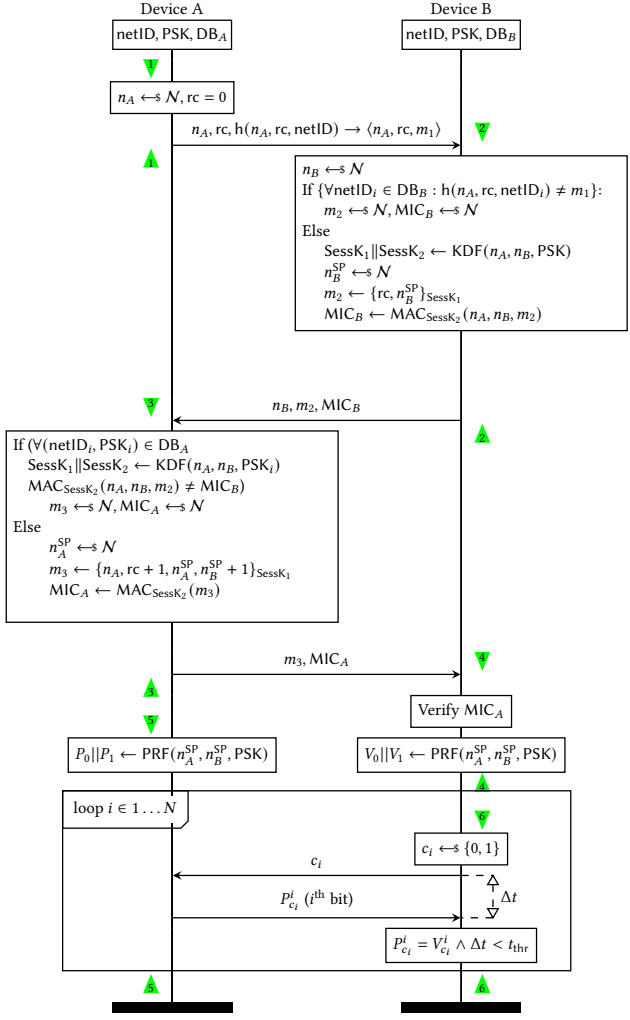

	\centering
\vspace{-6mm}
\scalebox{0.7}{
	\begin{msc}[instance distance=3.2cm, draw frame=none, msc keyword={},title top distance=0pt,title distance=0pt,
	first level height=2mm, level height=2mm]{}
\declinst{a}{Device A}{$\netID,\PSK, \DB_A$}
\declinst{b}{Device B}{$\netID,\PSK,\DB_B$}
\msccomment[msccomment distance=-4mm,side=left]{1}{a}%
\nextlevel[2]
\action*{$n_A\sample\mathcal{N},\rc=0$}{a}
\nextlevel[4]
\msccomment[msccomment distance=4mm,side=left]{2}{b}%
\nextlevel[2]
\mess{$n_A,\rc,\hash(n_A,\rc,\netID)\to\langle n_A,\rc,m_1\rangle$}{a}{b}
\msccomment[msccomment distance=-4mm,side=right]{1}{a}%
\nextlevel[1]
\action*{\begin{minipage}{6cm}
	$n_B\sample\mathcal{N}$\\
	If $\{\forall \netID_i\in\DB_B:\hash(n_A,\rc,\netID_i) \neq m_1\} $:\\
	\hspace*{3ex}$m_2 \sample \mathcal{N}, \MIC_B\sample\mathcal{N} $\\
	Else\\
	\hspace*{3ex}$\sessK_1\|\sessK_2\gets\KDF(n_A,n_B,\PSK)$\\
	\hspace*{3ex}$n_B^\mathsf{SP}\sample\mathcal{N}$\\
	\hspace*{3ex}$m_2\gets\{\rc,n_B^\mathsf{SP}\}_{\sessK_1}$\\
	\hspace*{3ex}$\MIC_B\gets\mathsf{MAC}_{\sessK_2}(n_A, n_B,m_2)$
	\end{minipage}}{b}
\nextlevel[23]
\msccomment[msccomment distance=-4mm,side=left]{3}{a}%
\nextlevel[2]
\mess{$n_B,m_2,\MIC_B$}{b}{a}
\msccomment[msccomment distance=4mm,side=right]{2}{b}%
\nextlevel[1]
\action*{\begin{minipage}{6cm}
    If ($\forall (\netID_i,\PSK_i)\in\DB_A$ \\
	\hspace*{2ex}$\sessK_1\|\sessK_2\gets\KDF(n_A,n_B,\PSK_i)$\\
    \hspace*{2ex}$\mathsf{MAC}_{\sessK_2}(n_A, n_B,m_2)\neq\MIC_B$)\\
	\hspace*{5ex}$m_3\sample\mathcal{N},\MIC_A\sample\mathcal{N}$\\
	Else\\
    \hspace*{5ex}$n_A^\mathsf{SP}\sample\mathcal{N}$\\
	\hspace*{5ex}$m_3\gets\{n_A,\rc+1,n_A^\mathsf{SP},n_B^\mathsf{SP}+1\}_{\sessK_1}$\\
	\hspace*{5ex}$\MIC_A\gets\mathsf{MAC}_{\sessK_2}(m_3)$\\
	\end{minipage}}{a}
	\nextlevel[21]
\msccomment[msccomment distance=4mm,side=left]{4}{b}%
\nextlevel[2]
\mess{$m_3,\MIC_A$}{a}{b}
\msccomment[msccomment distance=-4mm,side=right]{3}{a}%
\nextlevel[2]
\action*{Verify $\MIC_A$}{b}
\nextlevel[2]
\msccomment[msccomment distance=-4mm,side=left]{5}{a}%
\nextlevel[2]
\action*{$P_0||P_1\gets\mathsf{PRF}(n_A^\mathsf{SP},n_B^\mathsf{SP},\PSK)$}{a}
\action*{$V_0||V_1\gets\mathsf{PRF}(n_A^\mathsf{SP},n_B^\mathsf{SP},\PSK)$}{b}
\nextlevel[3]
\msccomment[msccomment distance=4mm,side=right]{4}{b}%
\nextlevel[2]
\inlinestart[right inline overlap=20.7mm,left inline overlap=20.7mm]{loop}{\colorbox{white}{loop $i\in 1\dots N$}}{a}{b}
\nextlevel[2]
\msccomment[msccomment distance=4mm,side=left]{6}{b}%
\nextlevel[2]
\action*{$c_i\sample\{0,1\}$}{b}
\nextlevel[4]
\measure[side=right]{$\Delta t$}{b}{b}[4]
\mess{$c_i$}{b}{a}
\nextlevel[4]
\mess{$P_{c_i}^i$ ($i$\textsuperscript{th} bit)}{a}{b}
\nextlevel
\action*{$P_{c_i}^{i}=V_{c_i}^{i}\land \Delta t<t_\mathsf{thr}$}{b}
\nextlevel[4]
\inlineend{loop}
\msccomment[msccomment distance=-4mm,side=right]{5}{a}%
\msccomment[msccomment distance=4mm,side=right]{6}{b}%
\nextlevel
\end{msc}
}

	\caption{Proposed \mbox{Wi-Fi} P2P persistent group formation procedure.}\label{fig:wifi-p2p-persistent-fix}
 \vspace{-0.1in}
\end{figure}

In our proposed \mbox{Wi-Fi} P2P persistent reconnection procedure, as shown in \autoref{fig:wifi-p2p-persistent-fix}, the revised 3-way handshake concludes with device $A$ sending at \encircleW{3} and device $B$ receiving at \encircleW{4}.Then a DiB check follows. Our method integrates the advertisement phase with a four-way handshake, initiating a series of strategic changes. Our first key modification in the advertisement phase is to hash the nonce $ n_A $, $ \mathsf{netID} $, and the replay counter $ rc $ like in the Simple Hash protocol: $ \mathsf{Hash}(n_A, \mathsf{rc}, \mathsf{netID}) $. Similarly, device $B$ resolves the $ \mathsf{netID} $ through the stored configuration file and responds with an encrypted replay counter, its own nonce, and a MIC. Note that the replay counter is encrypted using an ephemeral session key $ \mathsf{sessK} $. Following this, the response at \encircleW{3} encrypts both nonces and the incremented replay counter value, thereby concealing the relationship between the nonce and the replay counter values during the message exchanges at \encircleW{1}, \encircleW{2} and \encircleW{3}. As a result, the FO property is no longer violated because of the replay counter. \looseness=-1

Building on our Bluetooth solution, we introduce randomization for failure cases, making responses indistinguishable from those of successful condition outcomes. Specifically, at steps \encircleW{2}, \encircleW{3}, and \encircleW{4}, random values replace the messages of successful conditional checks. These values are sampled from a distribution that is indistinguishable from success messages, thereby hindering an adversary's ability to infer the outcome of the conditions. Consequently, if the adversary replays the initial message from $A$ at \encircleW{1}, condition-oblivious responses prevent the adversary from concluding the AInf attack. Should the adversary persist in scrutinizing later responses, a replay of the message at \encircleW{1} is detected at \encircleW{4}. Since the adversary does not have access to shared secrets, they are thwarted from replaying messages during subsequent phases of the protocol. The additional safey measure of proximity checks at \encircleW{5} and \encircleW{6} allows for the identification and neutralization of both relay and replay attacks. Device $B$ can then choose to silently ignore further communication attempts by the adversary.

\section{Tamarin Model and Evaluation}
\para{Tamarin Models} We use Tamarin-prover 1.6.1 on an Ubuntu 22.04 OS running on an Intel\textregistered{} i7-8550U CPU @ 1.80GHz × 8 processor with 16GB RAM. As shown in \autoref{tab:model-results}, we model the Bluetooth and \mbox{Wi-Fi} P2P reconnection procedures progressively. Starting from the advertisement phase models: \texttt{BT-AD}, \texttt{\mbox{Wi-Fi}-P2P-AD} and then extending the models with the session key derivation phase: \texttt{BT-AD-SKD}, \texttt{\mbox{Wi-Fi}-P2P-AD-SKD} for each case study.  This progressive modeling allows for a granular examination of privacy violations within distinct segments of the entire protocol. Our proposed protocol models are referred to as \texttt{BT-AD-SKD-PROPOSED} and \texttt{\mbox{Wi-Fi}-P2P-AD-SKD-PROPOSED}. These refined models validate all the targeted privacy properties. The code can be accessed in folder ``tamarin\_code'' in the provided artifact\footnote{\label{artifact-url}Anonymized for reviewing: \url{https://u.pcloud.link/publink/show?code=XgtitalK}}

Overall, our proposed Tamarin models comprise a total of 1,137 lines of code, along with approximately 20 security lemmata. The number of lemmata varies depending on the quantity of privacy-critical conditions identified in the protocols for each case study. \autoref{tab:model-results} further illustrates the proving run times relevant to the privacy properties of the proposed models. We observe that the FO models require the most substantial computational time, roughly 3 minutes for \texttt{BT-AD-SKD-PROPOSED} and \texttt{\mbox{Wi-Fi}-P2P-AD-SKD-PROPOSED}. This disparity, relative to other properties across both case studies, can be attributed to the complexity of verifying the diff-equivalence between two execution systems, necessitating a more extensive state space than the traditional trace-based properties. The models demonstrating vulnerabilities in both protocols also concluded within 3 minutes, revealing the failure of privacy properties.

\begin{table}[t]
\caption{Protocol models, security properties, and runtimes for the proposed models. Acronyms. \textbf{BT:} Bluetooth, \textbf{AD:} Advertisement, \textbf{SKD:} Session Key Derivation}
	\label{tab:model-results}
	\centering
 \scriptsize
	\setlength\tabcolsep{7pt}
	\begin{tabular}{l|c|c|c|r}
		\toprule[1.5pt]
		\textbf{Protocol}  &  \textbf{WA} &  \textbf{FO}  & \textbf{ND}  & \textbf{SP}	\\ \midrule[1.5pt]

		BT-AD      &            \xmark &             \xmark  &          \cmark &  \xmark     \\
		BT-AD-SKD              &            \xmark &              \xmark &          \cmark & \xmark     \\
		\midrule 
		BT-AD-SKD-PROPOSED              &            \cmark &              \cmark &          \cmark  & \cmark     \\
		(Runtimes)              &  (6s) &   (160s) & (1s)  &   (12s)    \\
				
		\midrule[1.5pt]
		\mbox{Wi-Fi}-P2P-AD              &            \xmark &              \xmark &          \cmark & \xmark     \\
		\mbox{Wi-Fi}-P2P-AD-SKD              &            \xmark &              \xmark &          \cmark  & \xmark     \\
		\midrule
		\mbox{Wi-Fi}-P2P-AD-SKD-PROPOSED              &            \cmark &              \cmark &          \cmark  & \cmark     \\
		(Runtimes)              &  (9s) &   (195s) & (2s)  &  (20s)    \\
		
		\bottomrule[1.5pt]
	\end{tabular}

\end{table}

\para{Performance Overhead} We developed a preliminary prototype to compute the performance overhead of our proposed design changes. The simplified protocol specifications -- as illustrated in Figures  \ref{fig:bleutooth-advertisement-n-SKD}, \ref{fig:bleutooth-advertisement-n-SKD-fix}, \ref{fig:wifi-p2p-persistent-simplified},
and \ref{fig:wifi-p2p-persistent-fix} -- are encoded as user-level C++ socket programs, and the messages are channeled through \mbox{Wi-Fi} and Bluetooth adapters between two Linux devices.
Since both of our machines do not have DiB measurement timing precision, we only implemented the challenge and response loops, passing one byte in each loop, without verifying the proximity through timing. The programs use 128-bit keys and nonces with AES provided by the Crypto++ library~\cite{cryptopp-library}. The responder device has an Intel i7-8550U CPU, 16GB RAM, Ubuntu 22.04 OS with an Intel Dual Band Wireless-AC 7265 wireless adapter, and an Intel Bluetooth wireless interface on HCI Version 4.2. The initiator device is equipped with an Intel i5-6200U CPU, 4GB RAM, Ubuntu 18.04.6 LTS with an Intel Wireless 3165 adapter, and the same Bluetooth configuration. The code can be accessed in folder ``cpp\_code'' in the provided artifact\footnotemark[1]. \looseness=-1

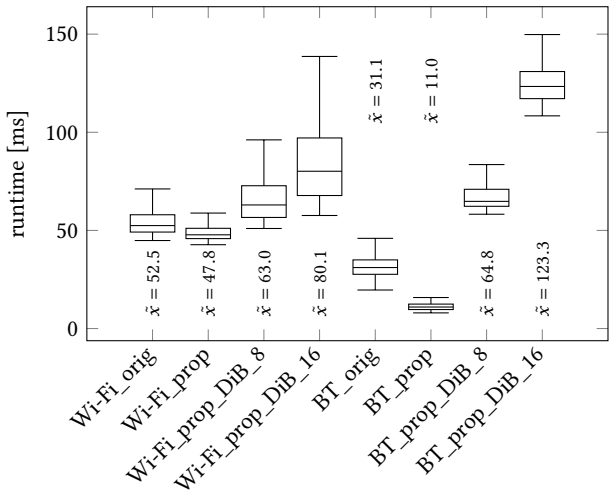
\begin{figure}[ht]
	\centering
		\begin{tikzpicture}
		 \begin{axis}[
			boxplot/draw direction=y,
			ylabel={runtime [ms]},
			height=6cm,
			width=\columnwidth,
			xtick={1,2,3,4,5,6,7,8},
    xticklabels={\mbox{Wi-Fi}\_orig,\mbox{Wi-Fi}\_prop,\mbox{Wi-Fi}\_prop\_DiB\_8,\mbox{Wi-Fi}\_prop\_DiB\_16,
    BT\_orig,BT\_prop,BT\_prop\_DiB\_8,BT\_prop\_DiB\_16
},
    xticklabel style = {rotate=45, anchor=east, yshift=-1mm},
		]

		\addplot+ [black, solid,boxplot prepared={
lower whisker = 44.845401 ,
lower quartile = 49.162264750000006 ,
median = 52.4575885 ,
upper quartile = 57.9794915 ,
upper whisker = 71.130845 ,
},] coordinates {};
\node at (axis cs: 1 , 2 ) [anchor=west, rotate=90] {\footnotesize $\tilde x =  52.5 $};
\addplot+ [black, solid,boxplot prepared={
lower whisker = 42.718755 ,
lower quartile = 45.8272 ,
median = 47.7608345 ,
upper quartile = 51.07248625 ,
upper whisker = 58.839156 ,
},] coordinates {};
\node at (axis cs: 2 , 2 ) [anchor=west, rotate=90] {\footnotesize $\tilde x =  47.8 $};
\addplot+ [black, solid,boxplot prepared={
lower whisker = 50.979823 ,
lower quartile = 56.60559825 ,
median = 62.9770305 ,
upper quartile = 72.76364099999999 ,
upper whisker = 96.130748 ,
},] coordinates {};
\node at (axis cs: 3 , 2 ) [anchor=west, rotate=90] {\footnotesize $\tilde x =  63.0 $};
\addplot+ [black, solid,boxplot prepared={
lower whisker = 57.598134 ,
lower quartile = 67.77739650000001 ,
median = 80.145374 ,
upper quartile = 97.105514 ,
upper whisker = 138.632254 ,
},] coordinates {};
\node at (axis cs: 4 , 2 ) [anchor=west, rotate=90] {\footnotesize $\tilde x =  80.1 $};

\addplot+ [black, solid,boxplot prepared={
lower whisker = 19.636 ,
lower quartile = 27.638761249999998 ,
median = 31.0760485 ,
upper quartile = 34.983242 ,
upper whisker = 45.960073 ,
},] coordinates {};
\node at (axis cs: 5 , 100 ) [anchor=west, rotate=90] {\footnotesize $\tilde x =  31.1 $};
\addplot+ [black, solid,boxplot prepared={
lower whisker = 7.986882 ,
lower quartile = 9.682061 ,
median = 10.967449 ,
upper quartile = 12.460485 ,
upper whisker = 15.809408 ,
},] coordinates {};
\node at (axis cs: 6 , 100 ) [anchor=west, rotate=90] {\footnotesize $\tilde x =  11.0 $};
\addplot+ [black, solid,boxplot prepared={
lower whisker = 58.243502 ,
lower quartile = 62.28418 ,
median = 64.7956235 ,
upper quartile = 70.95262925 ,
upper whisker = 83.542311 ,
},] coordinates {};
\node at (axis cs: 7 , 2 ) [anchor=west, rotate=90] {\footnotesize $\tilde x =  64.8 $};
\addplot+ [black, solid,boxplot prepared={
lower whisker = 108.370258 ,
lower quartile = 117.1393315 ,
median = 123.338727 ,
upper quartile = 130.941461 ,
upper whisker = 149.787619 ,
},] coordinates {};
\node at (axis cs: 8 , 2 ) [anchor=west, rotate=90] {\footnotesize $\tilde x =  123.3 $};

		 \end{axis}
		\end{tikzpicture}
		\vspace{-4mm}
		\label{fig:runtimes}

	\caption{Boxplot distribution of runtimes for the original specification (\_orig), proposed protocols without DiB (\_prop), and with DiB of 8 (\_DiB\_8) and 16 rounds (\_DiB\_16) of challenge and responses. Median runtimes are indicated with $\tilde x$ and the solid line in the box. Note that the standard outliers (1.5$\cdot$IQR below and above Q1 and Q3 respectively) values are omitted.}\label{fig:boxplots}
	\end{figure}

The runtimes for \mbox{Wi-Fi} and Bluetooth in \autoref{fig:boxplots} are box-plots of 500 runs. Each runtime spans wall clock time from the beginning of the advertisement until the end of the session key establishment phase/ distance bounding check phase, as observed by the initiator $A$.  We observe that our proposed changes ( \emph{prop} boxplot) in the initial three rounds (3-way handshake) of message sequences have equivalent/less runtime than the original protocol (\emph{orig} boxplot ). Only the round trips of the DiB challenge/response add significant overhead. The preliminary experiment suggests that as long as a distance bounding check can be optimized efficiently, the proposed protocol design can be implemented without a significant overhead.

\section{Discussion}
\label{sec:discussion}
\para{Limitations} Our research exhibits three salient limitations. First, the utilization of condition-oblivious responses, integral to the proposed solution, incurs an additional computational overhead. In the proposed protocols, condition-oblivious responses replace silent discard, which is more energy and performance-efficient. A potential compromise might allow users to toggle these features, adjusting the balance between privacy and performance according to individual needs. For instance, public figures or federal security personnel might willingly sacrifice some performance for more robust privacy protections. Second, allowlist groups for AInf can inadvertently misidentify target users, particularly in complex household scenarios where multiple family members access devices together. Such ambiguity can lead to false-positive identifications, misattributing a target user's actions to another family member. Likewise, non-member inference may occur as a consequence of packet errors caused by wireless signal interference, which can yield false negatives, potentially misidentifying a vulnerable victim as non-vulnerable. Our process algebra formalization describes the possible processes with allowlists, but misses a rigorous proof that our additional requirements (distance bounding) are sufficient. Lastly, our formal modeling is not fully automated, requiring significant manual intervention for building and debugging the Tamarin model.  \looseness=-1

\para{Future Work} In terms of future developments, First, we plan to redesign the Bluetooth and \mbox{Wi-Fi} reconnection procedures, mandating changes to the stack code including comprehensive consideration of inconspicuous details of channel changes in the metadata exchange. Regarding the formalization, we would like to prove the absence of Ainf attacks given our proposed countermeasures. This will hopefully enhance existing privacy tools to adopt novel techniques to capture broader classes of privacy attacks. \looseness=-1

\para{Ethics Consideration}
We responsibly disclosed our findings to both the Bluetooth SIG and \mbox{Wi-Fi} Alliance. Both groups mentioned that due to the growing importance of privacy, they want to prioritize privacy protection for ongoing and future development efforts. They acknowledged our discoveries and proposed countermeasures. Additionally, they plan to inform their stakeholders after our publication.

\section{Conclusion}

We have addressed a privacy concern, wherein an individual's association with sensitive groups can be inferred from wireless communications using allowlists. Our research clarifies how this privacy violation occurs, specifically within the context of \mbox{Wi-Fi} P2P persistent group formations and Bluetooth reconnection procedures. Through the development of formal models, we have not only validated the existence of this privacy violation but also devised a multifaceted strategy—including the use of oblivious and unpredictable responses, replay resistance, and proximity checks—to mitigate the risk. Our study enhances the understanding of privacy formalization and highlights critical privacy issues in prevalent wireless protocols. Overall, our work represents a significant step toward fostering privacy-preserving designs in the domain of wireless communication.


\bibliographystyle{ACM-Reference-Format} 

\bibliography{Bibliography/current,Bibliography/bluetooth-pairing-ndss-23,ble}

\end{document}